\documentstyle[11pt]{report}
\begin{document}

\renewcommand{\thefootnote}{\fnsymbol{footnote}}

\begin{titlepage}
\hfill MS-TPI-94-04 \\

\begin{centering}
\vfill
{\LARGE \bf Nonassociative Algebras and\\
            Nonperturbative Field Theory\\
         \vspace{2mm}
            for Hierarchical Models
            \footnote[1]{Work supported by the Deutsche
            Forschungsgemeinschaft
            under Grant Wi 1280/2-1}}

\renewcommand{\thefootnote}{\arabic{footnote}}

\vspace{2cm}
{\bf A. Pordt and C. Wieczerkowski} \\[6mm]

\vspace{0.3cm}
{\em
\, Institut f\"ur Theoretische Physik I, Universit\"at M\"unster, \\
   Wilhelm-Klemm-Str.\ 9, D-48149 M\"unster, Germany \\}

\vspace{2cm}
{\bf Abstract} \\
\end{centering}
\vspace{0.2cm}

Hierarchical renormalization group (RG) transformations are related to
nonassociative algebras.
These algebras serve as a new basic tool for
a rigorous treatment of global RG flows and the search of nontrivial
infrared fixed points. Convergent expansion methods are presented
and analyzed by the introduction of algebra-norms.
It is shown that the infrared fixed points can be investigated by solving
a quadratic equation with a finite number of unknowns.
A continuous manifold of two-dimensional periodic nontrivial fixed points
is given in terms of Theta-functions.
The local Borel-summability of the $\epsilon$-expansion for $l_*$-well
fixed points, $l_*\in \{ 2,3,\ldots ,\} $ is shown by
using algebraic methods.

\vspace{0.3cm}\noindent

\vfill \vfill
\noindent
MS-TPI-94-04\\
May 1994
\end{titlepage}


\newcommand{\nc}{\newcommand}
\nc{\be}{\begin{equation}}
\nc{\ee}{\end{equation}}
\nc{\bea}{\begin{eqnarray}}
\nc{\eea}{\end{eqnarray}}
\nc{\rbo}{\raisebox}
\nc{\cH}{{\cal H}}
\nc{\RR} {\rangle \! \rangle}
\nc{\LL} {\langle \! \langle}
\nc{\rmi}[1]{{\mbox{\small #1}}}
\nc{\eq}{eq.~}
\nc{\nr}[1]{(\ref{#1})}
\nc{\ul}{\underline}
\nc{\cM}{{\cal M}}
\nc{\mc}{\multicolumn}
\nc{\todo}[1]{\par\noindent{\bf $\rightarrow$ #1}}

\nc{\nonu}{\nonumber}
\nc{\lmax}{l_{\rm max}}
\nc{\half}{\mbox{\small$\frac12$}}
\nc{\eights}{\mbox{\small$\frac18$}}
\nc{\bg}{\bar \gamma}


\nc{\rg}{renormalization group }
\nc{\hrg}{hierarchical renormalization group }
\nc{\rgt}{renormalization group transformation }
\nc{\HHH}{{\cal H}}
\nc{\VVV}{{\cal V}}
\nc{\zzz}{{\cal Z}}
\nc{\rrr}{{\cal R}}
\nc{\lap}{(-\triangle)^{-1}}
\nc{\zeile}{\\[2mm]}

\newtheorem{lemma}{Lemma}[section]
\newtheorem{corollary}{Corollary}[section]
\newtheorem{theorem}{Theorem}[section]
\newtheorem{proposition}{Proposition}[section]
\newtheorem{definition}{Definition}[section]
\def\abstand{{\vspace{5mm}}}
\def\VVert{{|\! \! \; |\! \! \; | }}
\def\hga{{\widehat \gamma }}
\def\twtimes{{\widetilde \times }}
\def\lst{{l^*}}
\def\dst{{d^*}}
\def\bC{{\CCC}}
\def\bN{{\NNN}}
\def\bR{{\RRR}}
\def\bZ{{\ZZZ}}
\def\cB{{\cal B}}
\def\cC{{\cal C}}
\def\cF{{\cal F}}
\def\cL{{\cal L}}
\def\cN{{\cal N}}
\def\cP{{\cal P}}
\def\cR{{\cal R}}
\def\cS{{\cal S}}
\def\cT{{\cal T}}
\def\cU{{\cal U}}
\def\om{{\overline m}}
\def\oU{{\overline U}}
\def\g{{\gamma}}
\def\l{{\lambda}}
\def\L{{\Lambda}}
\def\rv{{\rm v}}
\def\eins{{\bf 1}}
\def\Gau#1#2{d\mu_{#1}({#2})}
\def\iGau#1#2{\int d\mu_{#1}({#2})}
\def\Gaug{d\mu_{\g}({\Phi })}
\def\Gaurv{d\mu_{\rv }({\Phi })}
\def\iGaug{\int d\mu_{\g}({\Phi })}
\def\GauN#1#2{d\mu_{#1}^{(N)}({#2})}
\def\iGauN#1#2{\int d\mu_{#1}^{(N)}({#2})}
\def\iGaugN{\int d\mu_{\g}^{(N)}({\Phi })}
\def\GaugN{d\mu_{\g}^{(N)}({\Phi })}
\def\GaurvN{d\mu_{\rv }^{(N)}({\Phi })}


\def\RRR{{\rm I\!R}}
\def\NNN{{\rm I\!N}}
 \def\one{{\mathchoice {\rm 1\mskip-4mu l} {\rm 1\mskip-4mu}
 {\rm 1\mskip-4.5mu l} {\rm 1\mskip-5mu l}}}
\def\QQQ{{\mathchoice
 {\setbox0=\hbox{$\displaystyle\rm Q$}\hbox{\raise 0.15\ht0\hbox to0pt
 {\kern0.4\wd0\vrule height0.8\ht0\hss}\box0}}
 {\setbox0=\hbox{$\textstyle\rm Q$}\hbox{\raise 0.15\ht0\hbox to0pt
 {\kern0.4\wd0\vrule height0.8\ht0\hss}\box0}}
 {\setbox0=\hbox{$\scriptstyle\rm Q$}\hbox{\raise 0.15\ht0\hbox to0pt
 {\kern0.4\wd0\vrule height0.7\ht0\hss}\box0}}
 {\setbox0=\hbox{$\scriptscriptstyle\rm Q$}
 \hbox{\raise 0.15\ht0\hbox to0pt
 {\kern0.4\wd0\vrule height0.7\ht0\hss}\box0}}}}
\def\CCC{{\mathchoice
 {\setbox0=\hbox{$\displaystyle\rm C$}\hbox{\hbox to0pt
 {\kern0.4\wd0\vrule height0.9\ht0\hss}\box0}}
 {\setbox0=\hbox{$\textstyle\rm C$}\hbox{\hbox to0pt
 {\kern0.4\wd0\vrule height0.9\ht0\hss}\box0}}
 {\setbox0=\hbox{$\scriptstyle\rm C$}\hbox{\hbox to0pt
 {\kern0.4\wd0\vrule height0.9\ht0\hss}\box0}}
 {\setbox0=\hbox{$\scriptscriptstyle\rm C$}\hbox{\hbox to0pt
 {\kern0.4\wd0\vrule height0.9\ht0\hss}\box0}}}}
%
\font\fivesans=cmr5
\font\sevensans=cmr7
\font\tensans=cmr10
\newfam\sansfam
\textfont\sansfam=\tensans\scriptfont\sansfam=
 \sevensans\scriptscriptfont
\sansfam=\fivesans
\def\sans{\fam\sansfam\tensans}
\def\ZZZ{{\mathchoice
 {\hbox{$\sans\textstyle Z\kern-0.4em Z$}}
 {\hbox{$\sans\textstyle Z\kern-0.4em Z$}}
 {\hbox{$\sans\scriptstyle Z\kern-0.3em Z$}}
 {\hbox{$\sans\scriptscriptstyle Z\kern-0.2em Z$}}}}

\thispagestyle{empty}
\tableofcontents
\label{SECintro}
In the \rg \cite{W71,WK74} approach one thinks of a Euclidean quantum
field theory in terms of its \rg flow.
Technically the idea is to split an infinite dimensional
path integral into finite dimensional portions
corresponding to fluctuations on increasing length scales
which can be performed stepwise one at a time.
A single step then defines a transformation of a
bare measure of function space into a renormalized one.
The most important problem in this approach is to
find the fixed points of this \rgt and to study the
flow in their vicinity.

In constructive quantum field theory
\cite{GJ87,R91} a rigorous version of
the \rg plays the r\^ole of an organizing principle of
convergent expansions. In particular polymer expansions
\cite{B84,GK71} can be generated from a \rg equation \cite{BY90}.
The existing technology is however limited to models
where one has a small expansion parameter at hand.
Examples are asymptotically free theories where the
expansion parameter is the running coupling constant
\cite{GK80}.
A number of important problems remain out of reach of
presently available methods. Among these are the infrared
behavior of the two dimensional $O(N)$-invariant nonlinear
$\sigma$-model \cite{GK86,PR91,I91}
and nonabelian gauge theories in four dimensions
\cite{B88a,B88b}.

The contribution of this report to the subject is a
formulation of the renormalization group in terms of
certain nonassociative algebras which will be called
\rg algebras. The idea is to define a bilinear composition
of measures on function space which equippes it with
the structure of a nonassociative algebra such that
a \rgt becomes an algebraic operation.
In particular the fixed point equation becomes algebraic
in terms of this composition.
We believe that a solution to the above problems can be
given in terms of a structural theory of these algebras.

Our general goal is to device algebraic methods
for nonperturbative studies of quantum fields.
Nonassociative products are also underlying the tree
formulas in the \rg approach of \cite{G85}.
In a nonassociative algebra there are different ways
of multiplying a given number of elements.
The different products are labelled by
trees. For our \rg algebras these trees correspond to
the ones of \cite{G85}.

In the case of hierarchical models this product is
a convolution composed with a dilatation.
The advantage of the algebraic point of view is
that it disentangles algebraic and combinatoric
aspects of a \rg analysis from analytic ones.
In this report we present a general theory of hierarchical
models based on nonassociative algebras.
The formulation in terms of polymer algebras
for the full models is under investigation.

\section{Hierarchical Models}
In this report we will restrict our attention to scalar
hierarchical models and discrete block spin transformations.
A generalization to $N$-component models is
directly possible. It will be omitted for the sake of
brevity. We expect also hierarchical gauge theories to be
suited for a similar treatment.

Hierarchical models have been investigated by many authors
\cite{B72,BS73,CE77,CE78,D69a,D69b,F87,KZ90,P93,PW91}
in various setups.
Here we will consider disrete hierarchical renormalization
group transformations for Euclidean scalar lattice fields
using the terminology of \cite{KW86,KW91}.
Let us recall the construction of hierarchical models.

We start from a Euclidean scalar lattice field
as a perturbation of a massless Gaussian measure.
Let $\Lambda=\ZZZ^d/L^N\ZZZ^d$ be a unit lattice
torus with side length $L^N$.
Let $\HHH$ be the space of real valued functions
$\Phi:\Lambda\rightarrow\RRR$ with mean zero and scalar
product $(\Phi,\Psi)=\sum_{x\in\Lambda}\Phi(x)\Psi(x)$.
Elements of $\HHH$ are called lattice fields.
A Euclidean scalar lattice field theory on $\Lambda$
is then defined by a massless Gaussian measure
${\rm d}\mu_{\lap}(\Phi)$ on $\HHH^\prime$ together
with a local interaction $V(\Phi)=\sum_{x\in\Lambda}
\VVV(\Phi(x))$ such that $Z(\Phi)=e^{-V(\Phi)}$ is
integrable with respect to ${\rm d}\mu_{\lap}(\Phi)$.
It is called polynomial if $\VVV(\Phi)$ is a polynomial.
The standard example is
\begin{equation}
\VVV(\Phi)=\kappa+\frac{\mu^2}{2} \Phi^2+\frac\lambda{4!} \Phi^4.
\end{equation}
The parameter $\kappa$ is called bare vacuum energy,
the parameters $\mu^2$ and $\lambda$ are
called bare coupling constants.
The massless covariance is given by its kernel
\begin{equation}
\lap(x,y)=\frac1{\vert\Lambda\vert}
\sum_{p\in\widetilde{\Lambda}\setminus\{0\}}
e^{ip(x-y)}\left(2\sum_{\mu=1}^d(1-{\rm cos}\, p_\mu)
\right)^{-1}.
\end{equation}
The problem is to construct and analyze the thermodynamic
limt $N\rightarrow\infty$ of correlation functions in
such a theory.
The block spin \rg is a tool to perform such an analysis.
A technical complication is the treatment of
nonlocal terms in the effective actions.
Hierarchical models are designed such that
the effective actions remain local.
They are obtained by replacing the massless covariance by
a hierarchical version
\begin{equation}
\lap_{{\rm hier}}(x,y)=
\sum_{M=0}^{N-1}L^{(2-d)M}\,\gamma\,
\delta_{[L^{-M}x],[L^{-M}y]}.
\end{equation}
Here $[L^{-M}x]$ denotes the integer part of $L^{-M}x$.
$\gamma$ is a positive parameter.
The full and the hierarchical model share a similar
critical behavior. However in the hierarchical model
translation invariance is manifestly broken and Osterwalder-Schrader
positivity does not hold.

Hierarchical models are ideally suited for the \rg approach.
Let us recall the definition of a hierarchical
block spin transformation.
Consider a second lattice
$\bar\Lambda=\ZZZ^d/L^{N-1}\ZZZ^d$
reduced in size by a factor of $L$.
This second lattice is a rescaled block lattice.
Let $\bar\HHH$ be the corresponding space of functions
$\bar\Phi:\bar\Lambda\rightarrow\RRR$.
Let $C:\HHH\rightarrow\bar\HHH$ be the linear operator
given by
$C\Phi(\bar x)=\sum_{x\in\Lambda}
C(\bar x, x) \Phi(x)$ with
\begin{equation}
C(\bar x, x) = \left\{
\begin{array}{ll}
L^{-d} & {\rm if }\quad\bar x^\mu\leq
[L^{-1}x^\mu]<\bar x^\mu+1, \\
0 & {\rm else}.
\end{array}
\right.
\end{equation}
$C$ is the block average operator. Let $C^T:\bar\HHH
\rightarrow\HHH$ be the transpose of $C$.

Let us rename the hierarchical massless covariance
$\lap_{{\rm hier}}$ on $\Lambda$ into $v$ and denote by
$\bar v$ the corresponding one on $\bar\Lambda$.
Then we have a splitting
\begin{equation}
v(x,y)=L^{2+d}\,(C^T\bar v C)(x,y)+\Gamma (x,y),
\end{equation}
where $\Gamma$ is given by
\begin{equation}
\Gamma (x,y)=\gamma\, \delta_{x,y}.
\end{equation}
We are led to consider the \rgt given by
\begin{equation}
\bar Z(\bar\Phi)=
\int {\rm d}\mu_{\Gamma}(\zeta)\,
Z(L^{1+\frac d2}\, C^T\bar\Phi+\zeta).
\end{equation}
The operator $\Gamma$ is called the fluctuation covariance.
Its most important property is that it is ultra local.
Therefore the \rgt preserves the locality of an interaction.
Suppose that $Z(\Phi)$ factorizes into a product
$\prod_{x\in\Lambda}\zzz(\Phi(x))$. Then so does
the effective interaction $\bar Z(\bar\Phi)$ into
$\prod_{\bar x\in\bar\Lambda}\bar\zzz(\bar\Phi(\bar x))$.
Note however that the effective interaction of a
polynomial theory is not anymore polynomial.
For a local theory the hierarchical \rgt is equivalent
to the nonlinear transformation
\begin{equation}
\bar\zzz (\Phi )=
\left( \int {\rm d}\mu_\gamma (\zeta )\,
\zzz (L^{1-\frac d2} \Phi+\zeta)
\right)^{L^d}
\end{equation}
of the local interaction Boltzmann factors. This transformation
is a transformation on functions of a single variable.
Remarkably it depends on the lattice geometry only through
the dimension parameter $d$ and the block length parameter $L$.
Both can be taken real valued extending hierarchical models to
intermediate dimensions. In the following we will let $L$
depend on $d$ through $L^d=2$. This corresponds to the case
when every block contains two lattice points.
In the algebraic formulation the fixed point equation
then becomes quadratic. The case when $L^d$ is an integer
larger than two also fits nicely into our scheme. In this
case the fixed point equation becomes an algebraic equation
involving higher powers.
The critical properties are expected to be independent of $L$
but do depend sensitively on the dimension $d$.
Let us further trade $d$ for $\beta=L^{1-\frac d2}$.
Substituting $\bar\zzz$ by $\bar\zzz^{L^d}$ we then obtain
\begin{equation}
\bar\zzz(\Phi)=
\int{\rm d}\mu_\gamma (\zeta)\,
\zzz(\beta\Phi+\zeta)^2.
\end{equation}
Let us then define a renormalization group operator $\rrr$
depending on two parameters $\beta$ and $\gamma$ by
\begin{equation}
\rrr_{\beta,\gamma}Z(\Phi)=
\int {\rm d}\mu_\gamma(\zeta)\,
Z(\beta \Phi+\zeta)^2.
\label{ren}
\end{equation}
This will be the final form of the hierarchical renormalization
group transformation here. Let us also introduce the notation
\begin{equation}
{\rm L}_X\rrr (Y)(\Phi)=
2\int {\rm d}\mu_\gamma (\zeta)\,
X(\beta\Phi+\zeta)\, Y(\beta\Phi+\zeta)
\end{equation}
for the linearization of $\rrr$ at $X$.

We will consider two different choices for $\beta$ and
$\gamma$. The first case is the one explained above
with $\beta=2^{\frac{2-d}{2d}}$. The second case is related
to the first by a transformation of the space of functions
factorizing out a quadratic fixed point.
This transformation leaves the form of the \rgt invariant
and only changes the values of the parameters. In the second
case $\beta$ is replaced by $L^{-2}\beta$ and $\gamma$ by
$L^{-2}\gamma$.
We will always assume that $d>2$ unless it is explicitely
stated differently.

\section{Ultraviolet Fixed Point}
One fixed point of eq.~(\ref{ren}) is immediately found.
It is simply
\begin{equation}
Z^{(0)}(\Phi)=1.
\end{equation}
It corresponds to the pure massless field.
This fixed point will be called ultraviolet fixed point.
The linearization of $\rrr$ at this trivial fixed point
is given by
\begin{equation}
{\rm L}_{Z^{(0)}}\rrr (Z)(\Phi)=
2\int {\rm d}\mu_\gamma(\zeta)\, Z(\beta\Phi+\zeta).
\end{equation}
It can be diagonalized exactly. The eigenfunctions are
normal ordered products
\begin{equation}
:\Phi^n:_{\gamma^\prime}=
\left(\frac{\gamma^\prime}2\right)^{\frac n2}
{\rm H}_n\left(\frac\Phi{\sqrt{2\gamma^\prime}}\right)
\end{equation}
given by rescaled Hermite polynomials. The normal
ordering covariance is given by
$\gamma^\prime=(1-\beta^2)^{-1}\gamma$.
The corresponding eigenvalues are
\begin{equation}
2^{\lambda^{(0)}_n}=
2^{1+n\frac{2-d}{2d}}.
\end{equation}
In the \rg language eigenfunctions with $\lambda^{(0)}_n>0$
are called relevant, those with $\lambda^{(0)}_n=0$ marginal,
and the others irrelevant.

The number of relevant eigenfunctions depends on $d$.
Let us restrict our attention to even eigenfunctions.
One observes a sequence of critical dimensions
$d_2>d_3>d_4>\dots>2$ given by
\begin{equation}
d_{n}=\frac{2n}{n-1}
\end{equation}
Below the critical dimension $d_{n}$ the eigenfunction
$:\Phi^{2n}:_{\gamma^\prime}$ becomes relevant
whereas it is irrelevant above $d_{n}$.
In particular above $d=4$ only $1$ ($n=0$) and
$:\Phi^2:_{\gamma^\prime}$ ($n=1$) are relevant.

We think of the subspace spanned by the relevant
eigenfunctions as a tangent space to an unstable
manifold at the fixed point. Imagining normal coordinates
points in a neighbourhood of the fixed point, points of
the tangent space can be thought of as points
on the manifold itself.
The dimension of the unstable manifold of the
ultraviolet fixed point changes as the dimension is
lowered from the region above four.

\section{High Temperature Fixed Point}
A second fixed point is not hard to find either. It is given
by
\begin{equation}
Z^{(1)}(\Phi)=e^{-V^{(1)}(\Phi)}
\end{equation}
with
\begin{equation}
V^{(1)}(\Phi)=\frac a2 \Phi^2+b,
\end{equation}
where $a=(L^2-1)/{\gamma L^d}$. It is called high temperature
fixed point. The linearized transformation at the high
temperature fixed point can also be diagonalized exactly.
The eigenfunctions are given by
\begin{equation}
:\Phi^n:_{\gamma^{\prime\prime}} \, Z^{(1)}(\Phi)
\end{equation}
with
$\gamma^{\prime\prime}=(1-L^{-2}\beta^2)^{-1}L^{-2}\gamma$.
The corresponding eigenvalues are
\begin{equation}
L^{\lambda_n^{(1)}}=2^{1-n\frac{2+d}{2d}}.
\end{equation}
Let us again consider even eigenfunctions only.
We see that all eigenfunctions with $n> 1$ are irrelevant
for the whole range $d>2$ we are considering here.
In this sense the high temperature fixed point is stable.

Let us remark that at any fixed point the fixed point itself is
a relevant eigenfunction with eigenvalue $\lambda=1$.
This eigenfunction is called the volume eigenfunction.
A perturbation by it does not change physical properties
since it corresponds to a trivial rescaling.

For $d\leq 4$ there are no further fixed point given by
functions $Z(\Phi)$ which are integrable with respect
to the Gaussian measure ${\rm d}\mu_\gamma(\Phi)$.

Not all integrable measures are however driven into
the mentioned two fixed points upon iteration
of renormalization group transformations.
Noncritical theories in the unbroken phase are
attracted by the high temperature fixed point. Those
in the broken phase approach a singular fixed point in
form of an infinitely deep double well.
This observation suggests to widen the idea of fixed
points to a broader class in order to incorporate
fixed points which only exist asymptotically or do not
correspond to integrable functions.
We will not pursue this line of thoughts here but would
like to mention that the notion of nonexistence of
other fixed points refers to a certain class of
admissible functions.

\section{Nontrivial Fixed Points}
For $d\geq 4$ no further fixed points exist besides the
ultraviolet and the high temperature fixed point.
As the dimension is lowered new fixed points appear
sequentially at the critical dimensions
$d_{n}$.

The first threshold is $d_2=4$. Below four dimensions
a nontrivial fixed point $Z^{(2)}(\Phi)$ with
double well shaped potential $V^{(2)}(\Phi)$ exists.
This fixed point is called infrared fixed point.
At $d=3$ it has been rigorously constructed in \cite{KW86,KW91}.

The next threshold is $d_3=3$ at which $Z^{(3)}(\Phi)$
appears and so on. Below $d_{n}$ a fixed point
$Z^{(n)}(\Phi)$ exists. The potential $V^{(n)}(\Phi)$
has the shape of an $n$-well.
The linearized transformation at $Z^{(n)}(\Phi)$
has $n$ relevant eigenfunctions including a trivial
volume eigenfunction. In particular the
infrared fixed point has two relevant eigenfunctions.
They have not been calculated exactly.
Numerical results can be found in \cite{PPW94}.
The corresponding
eigenvalues are related to critical exponents.
The nontrivial fixed point $Z^{(n)}(\Phi)$
approaches $Z^{(0)}(\Phi)$ as $d$ goes to $d_{n}$
from below. In this sense all nontrivial fixed points
bifurcate from the ultraviolet fixed point as the
dimension parameter is varied.

An intersting question is that of critical lines inbetween
fixed points. The correlation length at a fixed point is
either zero or infinite. A fixed point is called critical
if the correlation length is infinite. The ultraviolet
fixed point and all nontrivial fixed points are critical
whereas the high temperature fixed point is noncritical.
The critical theories form a submanifold within the set
of all theories which contains the critical fixed points.
On this critical manifolds the \rg flow always ends in
a critical fixed point. Consider for instance the situation
inbetween three and four dimensions where there are only
two critical fixed points. The two fixed points are
connected by a critical line of critical theories.
Tangent to this line at the ultraviolet fixed point is
the relevant eigenfunction $:\Phi^4:_{\gamma^\prime}$.
The flow on this critical line goes from the ultraviolet
to the infrared fixed point.
In general for $d_{n+1}\leq d\leq d_{n}$ there
originate critical lines from $Z^{(0)}(\Phi)$ to all
fixed points $Z^{(n)}(\Phi),\dots,Z^{(2)}(\Phi)$
corresponding to the different relevant eigenfunctions of
the linearized transformation at the ultraviolet fixed
point. Note that there is also a line from $Z^{(0)}(\Phi)$
to $Z^{(1)}(\Phi)$ but the theories on this line are
noncritical.
There are further critical lines from $Z^{(n)}(\Phi)$ to
$Z^{(n-1)}(\Phi),\dots,Z^{(2)}(\Phi)$ corresponding to
the unstable directions at the nontrivial fixed point
$Z^{(n)}(\Phi)$. Recall that the dimension of the
unstable manifold of $Z^{(n)}(\Phi)$ is $n$-dimensional.
Excluding one direction corresponding to the volume
eigenfunction and one yielding a line of noncritical
theories leading to the high temperature fixed point
we obtain a consistent picture.

A rigorous construction of all nontrivial fixed points has
only been given within the infinitesimal version of
the hierarchical \rg in the article of \cite{F87}.
A construction for the disrete case in the sense of
\cite{KW86,KW91} is still missing. The question of critical
lines has so far only been investigated using
heuristic methods.
Most prominent among these is the $\epsilon$-expansion.
The nontrivial fixed point $Z^{(n)}(\Phi)$ and the
linearization around it can be studied by
$\epsilon$-expansion at $d=d_{n}-\epsilon$.
We will show in detail how this $\epsilon$-expansion
can be done in the discrete hierarchical case.

\section{Renormalization Group Algebra}
For two functions $X(\Phi)$ and $Y(\Phi)$ we define a
bilinear composition depending on two parameters
$\beta$ and $\gamma$ by
\begin{equation} \label{betpro}
:X \times_{\beta,\gamma}Y:_\g (\Phi)=
\int{\rm d}\mu_{\gamma (1-\beta^2)}(\zeta)\,
:X:_\g (\beta\Phi+\zeta) \,:Y:_\g (\beta\Phi+\zeta).
\end{equation}
This composition is commutative but not associative if $\beta \ne 1.$
It becomes associative when $\beta=1$.
We call this operation the renormalization group product.
In terms of it the fixed point equation takes the
form
\begin{equation}
Z\times_{\beta,\gamma}Z(\Phi)=Z(\Phi).
\end{equation}
Let us expand $X$ in terms of monomials
\begin{equation}
X(\Phi)=\sum_{n=0}^\infty
X_n\, \Phi^{2n}.
\end{equation}
Again we consider only even functions.
In terms of the expansion coefficients we find
\begin{equation}
(X\times_{\beta,\gamma}Y)_l=
\sum_{n=0}^\infty\sum_{m=0}^\infty
\beta^{2l}\, \gamma^{n+m-l}\, \cC_l^{mn}\,
X_m\, Y_n.
\end{equation}
The $\cC_l^{mn}$ are integer valued coefficients
given by a decomposition formula for normal ordered products.
This equation expresses the \rg product in the basis
given by normal ordered products.
We see that the dependence on $\gamma$ is trivial.
It can be absorbed introducing
\begin{equation}
\bar X_n=\gamma^n\, X_n.
\end{equation}
In terms of the rescaled coefficients we find
\begin{equation}
(\bar X\times_{\beta}\bar Y)_l=
\sum_{n=0}^\infty\sum_{m=0}^\infty
\beta^{2l}\, \cC_l^{mn}\,
\bar X_m\, \bar Y_n.
\end{equation}
The \rg product will be analyzed in this form below.
In particular the fixed point equation becomes a
quadratic equation involving infinitely many variables
given by
\begin{equation}
(\bar Z\times_{\beta} \bar Z)_l=
\sum_{n=0}^\infty \sum_{m=0}^\infty
\beta^{2l}\, \cC_l^{mn}\,
\bar Z_m\, \bar Z_n.
\end{equation}
It will turn out that this representation is well adapted
to an algebraic study of fixed points and linearizations
around them. Let us remark that the product eq.~(\ref{betpro})
becomes associative when $\beta =1.$ But the covariance
on the rhs of eq.~(\ref{betpro}) $\g (1-\beta^2)$ becomes zero
in this case.

\section{Norms and Renormalization Group Algebra}
We will consider a number of different norms in our analysis.
Norms serve several purposes.
Consider for instance an approximate study of the \rg flow of
a hierarchical model. At the beginning we have to
truncate the system to a finite number of degrees of freedom.
The result will only be a good approximation if we can show
that the neglected remainders are small. This is most
efficiently done in terms of norm estimates.
In some sense the idea is to reduce the treatment of large fields
to norm estimates.

Norm estimates also give information about invariant subspaces
under \rg transformations. For instance with a suitable norm it
can be shown that the flow converges to zero if the initial
interaction is sufficiently small.

A norm $\Vert \cdot\Vert$ is called an algebra norm if
\begin{equation}
\Vert X\times Y\Vert\leq\Vert X\Vert\,\Vert Y\Vert .
\end{equation}
We immediately find that a fixed point has to satisfy
$\Vert Z\Vert\geq 1$.

Let us remark that a linear function $\omega$ from the space
of effective interactions to $\CCC$ is called a weight if it
has the property that $\omega(X\times Y)=\omega(X) \, \omega(Y)$.
In this case fixed points would obey $\omega(Z)=1$ and
the subspace with unit weight would be invariant.
Unfortunately, weights only exist for $\beta=1$.
For certain classes of nonassociative algebras in connection
with genetic bases weights play an important r\^ole \cite{B66}.

Norms replace weights in the nonassociative case.
Weights are nevertheless useful since they serve to construct
certain prenorms in our case.

An algebra together with an algebra norm defines upon completion
a Banach algebra. Banach algebras are very well studied objects
in mathematices in the associative case.

A nice application of norm estimates is the determination of
the large field behavior of the potential at the
infrared fixed point \cite{KW91}.
A new application that will be given below is a proof of
local Borel summability of the $\epsilon$-expansions for the
nontrivial fixed points.

\section{$\epsilon$-Expansion}
The $\epsilon$-expansion is as old as the \rg \cite{WK74}.
We will present a version for the hierarchical case which does not
rely on perturbation theory, see also \cite{CE77}. The idea is to expand
\begin{equation}
Z^{(n)}(\Phi)=\sum_{k=0}^\infty
Z^{(n)}_k(\Phi) \, \epsilon^k
\end{equation}
at $d=d_n-\epsilon$. Numerical studies show that low order
approximations already give the fixed points to a reasonable
accuracy \cite{ZJ89,PPW94}.
In terms of the representation by normal ordered products
the $\epsilon$-expansion will be written into
recursive form which can be evaluated using computer algebra.
The $\epsilon$-expansion for $Z^{(n)}(\Phi)$ does not
converge. It will however be shown to be locally Borel summable.

It is conceivable that at least in the hierarchical case
an $\epsilon$-expansion with remainder can be turned into
a rigorous construction of fixed points. Again remainder terms
will have to be estimated using norms.

\section{Critical Dimensions and
Renormalization Group Algebra}
The critical dimensions have been introduced as thresholds
where eigenfunctions of the linearized \rg transformation
at the ultraviolet fixed point become relevant.
They coincide with the critical dimensions at which new
fixed points bifurcate from the ultraviolet fixed point.
This can be seen by the following reasoning.
Consider the fixed point equation
$Z\times_\beta Z=Z$ and put $Z=Z_0+H$ with $Z_0$ the high
temperature fixed point. For $Z$ to be fixed point the
remainder has to satisfy $(I-2Z_0)\times_\beta H=H\times_\beta H$.
Here $I$ is a unit element adjoined to the algebra.
Treating $H$ as a perturbation of $Z_0$ we obtain
$H\in{\rm ker}((I-2Z_0)\times_\beta)$ as first order condition.
But in the basis given by normal ordered products we have
$((I-2Z_0)\times_\beta)={\rm diag}(1-2\beta^{2l})_{l=0,1,\dots}$.
Thus the kernel is nontrivial iff there is an $l$ such that
$1-\beta^{2l}=0$. This is equivalent to
$d=\frac{2l}{l-1}$.

\chapter{Introduction to Nonassociative Algebras}
\label{IntNonAss}

\section{Definition of the Nonassociative Algebra}
\label{DefNonAss}
We recall in this section some basic definitions for nonassociative
algebras \cite{B66,S66}.

Let $\cB $ be an infinite-dimensional vector space over $\bC $
with basis $e_0,e_1,e_2,\ldots $.
Let a multiplication $\times $ on $\cB $ be defined by {\em constants
of multiplication} $B_l^{mn}\in \bC $ for $l,m,n\in \bN $
and the table of multiplication
\be
e_m \times e_n = \sum_l B_l^{mn} \, e_l.
\ee
We suppose that the $\times $-multiplication obeys the distributive laws
\bea
a\times (b+c) &=& a\times b + a\times c
\nonu\\
(a+b)\times c &=& a\times c + b\times c
\eea
and
\be
(\l a)\times b = a\times (\l b) = \l a\times b
\ee
for all $a,b,c \in \cB $ and $\l \in \bC $.

We call an element $a\in \cB $ {\em finite of degree $n$} if $a_l=0$
for all $l\ne n.$ Let $\cB_n$ be the subspace of $\cB $ consisting
of elements of degree $n.$ The algebra $\cB $ is called {\em graded}
if
\be
\cB_m \times \cB_n \subseteq \cB_{m+n}.
\ee
The {\em direct sum} $\bigoplus_{n=0}^\infty \cB_n$ is the subalgebra of
$\cB $ consisting of sums of elements of finite degree.
In this chapter we will only
consider $\bigoplus_{n=0}^\infty \cB_n$. Later we will also consider closures
of $\bigoplus_{n=0}^\infty \cB_n$ with respect to certain norms.

Consider two elements
$a,b\in \cB $ defined by
\be
a := \sum_{m:\, m\ge 0} a_m e_m , \qquad b := \sum_{n:\, n\ge 0} b_n e_n .
\ee
Then, we have
\be
a\times b = \sum_{l:\, l\ge 0} c_l e_l,
\ee
where
\be
c_l = \sum_{m,n:\, m,n\ge 0} B_l^{mn} a_m b_n .
\ee
The vector space $\cB $ together with the multiplication $\times $
is called an {\em algebra } $(\cB ,\times )$ (with basis $e_0,e_1,e_2,
\ldots ).$ We call the algebra $(\cB ,\times )$ {\em commutative} iff
\be
a\times b = b\times a
\ee
for all $a,b\in \cB .$ This is equivalent to the following condition
for the constants of multiplication
\be
B_l^{mn} = B_l^{nm}
\ee
for all $l,m,n\in \bN $.
We call the algebra $(\cB ,\times )$ {\em associative} iff
\be
a\times (b\times c) = (a\times b)\times c,
\ee
for all $a,b,c\in \cB .$
This is equivalent to the following condition
for the constants of multiplication
\be
\sum_{l:\, l\ge 0} B_l^{mn}\, B_k^{lq} = \sum_{l:\, l\ge 0}
                                 B_k^{ml}\, B_l^{nq}.
\ee
We consider now a change of the basis of the algebra $\cB .$
Let $f_0, f_1 ,f_2, \ldots $ be another basis of the vector space $\cB $.
Suppose that
\be
e_n = \sum_{l=0}^\infty U_{nl} f_l
\ee
and that the matrix $U$ is invertible, i.~e.~, $U^{-1}$ exists, such that
\be
\sum_{l=0}^\infty U_{nl} (U^{-1})_{lk} = \delta_{nk} =
\sum_{l=0}^\infty (U^{-1})_{nl} U_{lk}.
\ee
Then the table of multiplication reads
\be
f_m \times f_n = \sum_l C^{mn}_l f_l ,
\ee
where the constants of multiplication are
\be \label{isocmnl}
C^{mn}_l = \sum_{m^{'},n^{'},l^{'}} (U^{-1})_{mm^{'}} (U^{-1})_{nn^{'}}
 U_{ll^{'}} B_l^{mn} .
\ee
Below we will consider examples where the transformation matrix $U$
is diagonal, i.~e.,
\be
U_{nl} = \l_n\, \delta_{nl}.
\ee
Then the constants of multiplication are related by
\be
C_l^{mn} = \frac{\l_l}{\l_m\, \l_n}\, B_l^{mn}.
\ee
Define a new multiplication $\twtimes $ for the basis elements $e_n$ by
\be
e_m \twtimes e_n = \sum_l C_l^{mn} e_l.
\ee
Then eq.~(\ref{isocmnl}) implies
\be
U(a\times b) = U(a) \twtimes U(b),
\ee
for all $a,b\in \cB $, i.~e.~, $U:\, (\cB ,\times ) \rightarrow
(\cB ,\twtimes )$ is an algebra-isomorphism.

For $a=\sum_{m=0}^\infty a_m e_m$, $b=\sum_{n=0}^\infty b_n e_n \in \cB $
define the canonical scalar product
\be
<a,b> = \sum_{n=0}^\infty a_n b_n.
\ee
Consider the function $F:\, \cB \bigotimes \cB \bigotimes \cB \rightarrow
\bR $ defined by
\be
F(a\otimes b\otimes c) = <a,b\times c>.
\ee
We call the algebra {\em symmetric} iff $F$ is a symmetric function.
The algebra is symmetric if the constant of multiplication
$B_l^{mn}$ is symmetric under permutations of $l,m$ and $n.$

In the following we will introduce the notion of weight and genetic
basis for the algebra $\cB $ (cf.~\cite{B66}).
We call a linear mapping $\omega :\, \cB \rightarrow \bC $ a {\em
weight} iff $\omega \ne 0$ and $\omega (u\times v) = \omega (u)
\omega (v)$ for all $u,v\in \cB .$ If there exists a weight $\omega $
for the algebra $\cB $ we call $\cB $ a {\em weighted algebra.}
We can easily see that $\omega (\cB )= \bC $, i.~e.~, $\omega $ is
surjective. Therefore
\be
\dim (\cB -\ker \omega ) =1 .
\ee
$\omega $ is a one-dimensional representation of the algebra $\cB .$
Let $\{ e_m,\ m\in \bN \} $ be a basis of $\cB .$ We call this basis
{\em genetic} iff
\be \label{genbas}
\sum_l B_l^{mn} =1,
\ee
for all $m,n\in \bN .$ $B_l^{mn}$ are the constants of multiplication
with respect to the basis $\{ e_m,\ m\in \bN \} .$
\begin{lemma} \label{lweigen}
Let $\cB $ be an algebra. Then there exists a genetic basis
$\{ e_m,\ m\in \bN \} $ iff there exists a weight $\omega $ of $\cB .$
\end{lemma}
\underline{Proof:\,} ``$\Rightarrow $'': Suppose the there exists a genetic
basis $\{ e_m,\ m\in \bN \} $ of $\cB $. Consider $a=\sum_{m=1}^N a_m e_m$,
$u_m\in \bC .$ Define
\be
\omega (a) := \sum_{m=1}^N a_m .
\ee
Obviously $\omega :\, \cB \rightarrow \bC $ is linear and non-zero.
We have, using eq.~(\ref{genbas}),
\be
\omega (u\times v) = \sum_{l,m,n} B_l^{mn} \, u_m v_n = \sum_{m,n} u_m v_m
= \omega (u)\, \omega (v).
\ee
Thus, $\omega $ is a weight.

\noindent
``$\Leftarrow $'': Suppose that there exists a weight $\omega :\,
\cB \rightarrow \bC .$ Since $\cB = (\cB - \ker \omega ) \oplus
\ker \omega $ and $\dim (\cB -\ker \omega )=1$ there exists a basis
$\{ u_m,\ m\in \bN \} $ such that $u_0 \in \cB -\ker \omega $,
$\omega (u_0)=1$, and $u_1,u_2,\ldots \in \ker \omega .$ Define a new basis
$\{ e_m, m\in \bN \} $ by
\be
e_0 := u_0,\qquad e_n:= u_0 + u_n,
\ee
for all $n\ge 1.$ Then, we have
\be
\omega (e_m\times e_n) = \omega (e_m) \, \omega (e_n) =1 .
\ee
Since
\be
\omega (e_m \times e_n ) = \sum_l B_l^{mn},
\ee
we see that eq.~(\ref{genbas}) holds. Therefore, the basis $\{ e_m,\
m\in \bN \} $ is genetic. $\  \  \Box $

\abstand
For two subsets $U,V\subseteq \cB $ define a subset $U\times V$ of $\cB $
by
\be
U\times V := \{ u\times v|\, u\in U,\  v\in V\} .
\ee
We call a subspace $U$ of the vector space $\cB $ a {\em subalgebra of
$\cB $} iff $U\times U\subseteq U$. Obviously, $(U,\times )$ is itself an
algebra, where the $\times $-multiplication is restricted to $U$.
We call a subspace $I$ of the vector space $\cB $ a {\em left (right)
ideal} of $\cB $ iff $\cB \times I \subseteq I (I\times \cB \subseteq
I).$ Let us remark that in the RG context the subalgebras (ideals)
play the r\^ole of invariant subspaces with respect to the (linearized)
RG transformation.

If $\cB $ does not contain a unit element $\eins $ we may adjunct an
unit element $\eins $ by
\be \label{einsad}
(\alpha \eins + a)\times (\beta \eins +b) = (\alpha \beta )\, \eins +
\alpha b + \beta a + a\times b,
\ee
where $\alpha ,\beta \in \bC $ and $a,b \in \cB $.


Let $\cB \bigotimes \cB $ be the tensor product. A multiplication on
$\cB \bigotimes \cB $ can be defined by
\be
(a\otimes b) \times (c\otimes d) = (a\times c) \otimes (b\times d).
\ee
Let us denote the basis elements of the tensor product $e_m \otimes
e_{m^{'}}$ by $e_{mm^{'}}$. Then the constants of mutiplication
$B^{mm^{'},nn^{'}}_{ll^{'}}$ of the tensor-algebra $B\bigotimes B$
are given by the following table of multiplication
\be
e_{mm^{'}} \times e_{nn^{'}} = \sum_{l,l^{'}} B^{mm^{'},nn^{'}}_{ll^{'}}
  e_{ll^{'}}
\ee
or explicitly
\be
B^{mm^{'},nn^{'}}_{ll^{'}} = B^{mn}_l B^{m^{'}n^{'}}_{l^{'}}.
\ee

Let $\Vert \cdot \Vert$ be a norm on $\cB $. We call $\Vert \cdot \Vert$
an algebra-norm on $(\cB ,\times )$ iff
\be
\Vert a\times b\Vert \le \Vert a \Vert \cdot \Vert b \Vert .
\ee
Let us remark that if $U:\, (\cB ,\times ) \rightarrow (\cB ,\twtimes )$
is an algebra-isomorphism defined by $U(a)=\l a$, $\l \in \bC $ and
$\Vert \cdot \Vert $ is a norm on $\cB $ obeying
\be
\Vert a\times b\Vert \le \l \, \Vert a \Vert \cdot \Vert b \Vert .
\ee
we have
\be
\Vert a\twtimes b\Vert \le \Vert a \Vert \cdot \Vert b \Vert ,
\ee
i.~e.~, $\Vert \cdot \Vert $ is an algebra-norm on $(\cB ,\twtimes ).$

A {\em Banach-algebra} is an algebra $(\cB ,\times )$ with algebra-norm
$\Vert \cdot \Vert $ such that the normed space $(\cB ,\Vert \cdot \Vert )$
is complete, i.~e.~, every Cauchy sequence converges.

We will now introduce the notion of regular quasi-representation for
nonassociative algebras. The regular quasi-representation for
nonassociative algebras correspond to the linearized RG transformations.
For an algebra $(\cB ,\times )$ and $a\in \cB $ define the linear operator
$L_\times (a):\, \cB \rightarrow \cB $ by
\be
L_\times (a)(b) = a\times b.
\ee
The matrix elements are given by
\be
L_\times (a)_{ln} := \sum_m B_l^{ml} \, a_m.
\ee
We omit the index $\times $ of $L_\times $ if there arises no ambiguities.
For nonassociative algebras we call $L_\times $ a {\em regular
quasi-representation.} Let us remark that $L_\times $ is not a
representation in the nonassociative case.
Supposing that $(\cB ,\times )$ is an associative algebra we may conclude
\be \label{linrep}
L(a\times b) = L(a)\circ L(b) .
\ee
Thus $L:\, (\cB ,\times )\rightarrow (Lin(\cB ,\cB ) ,\circ )$ is a
representation of the algebra $(\cB ,\times )$ which is called
{\em regular representation of} $(\cB ,\times )$. We will use the notation
$L(a) = (a\times )$. If the algebra is associative and commutative,
we have
\be
[L(a),L(b)] = 0,
\ee
for all $a,b\in \cB $. If $L(a)$ and $L(b)$ commute they have the common
eigenvectors. We call $c\ne 0$ an eigenvector of $L(a)$ if there exists
$\l (a,c) \in \bC $ such that
\be
L(a)\, c = \l (a,c) \, c .
\ee
Using relation (\ref{linrep}) we obtain
\be
\l (a\times b,c) = \l (a,c)\, \l (b,c).
\ee
Thus $\l (\cdot ,c):\, (\cB ,\times ) \rightarrow (\bC ,\cdot )$ is a
1-dimensional representation of $(\cB ,\times )$ for any eigenvector $c$.
Let us remark that for an idempotent element $z\in \cB $, i.~e.~, obeying
$z\times z=z$, we have $\l (z,z)=1$. This equation holds also for the
nonassociative case, i.~e.~, $z$ is an eigenvector for $L(z)$ with
eigenvalue 1.

In this paper we are concerned with algebras where the commutative
and associative multiplication $\times_1$ is deformed by a linear operator.
Define a $\times_h$-multiplication by
\be
a\times_h b = \cS_h (a\times_1 b),
\ee
for all $a,b\in \cB $.
$\cS_h:\, \cB \rightarrow \cB $ is a linear
operator defined by
\be
\cS_h(e_m) := h(m) \, e_m,
\ee
where $h:\, \bN \rightarrow \bC $. The cases we are mostly interested in are
$h(m)=\lambda^m $ and $h(m)=\lambda^{m^2}$.
This can be viewed as a 1-parameter deformation of the algebra $\cB $
giving up the associativity axiom.

Furthermore, suppose that $(\cB ,\times )$ is isomorphic to a
symmetric algebra. We call such algebras {\em renormalization group
algebras}. For $h\ne \eins $ the algebra is commutative and nonassociative.
If the constants of multiplication of $(\cB ,\times_1)$ are given by
$B_l^{mn}$ the constants of multiplication of $(\cB ,\times )$ are given
by $h(l)\, B_l^{mn}$. Supposing that $e_0$ is the unit element
of $(\cB ,\times_1)$
we see that $B_l^{0n} =\delta_{ln}$. This is equivalent to
$L_{\times_1}(e_0) =\eins $. Thus the eigenvalues of $L_{\times_1}$ are
equal to 1 for all $a\in B.$ Obviously,
\be
L_\times (e_0) = diag\, (h(0),h(1),h(2),\ldots ).
\ee
The vector space $\cB $ can be identified by its dual space $\cB^{'}$ if we
define a dual basis $e_m^{'}$ of $\cB^{'}$ by
\be
e_m^{'}(e_n) := \delta_{mn}.
\ee
Then we see that $L(\cB ) \subseteq \cB \bigotimes \cB $ and
\be
L(a) = \sum_{l,m,n} B_l^{mn}\, a_m\, e_l \otimes e_n.
\ee
In the following we are interested in commutative but nonassociative
algebras. For $x\in \cB $ the subalgebra $[x]$ generated by $x$ consists
of all products of $x$. If this subalgebra $[x]$ is associative for
all $x\in \cB $ then $\cB $ is called a {\em power algebra}. Unfortunately,
nonassociative algebras defined by RG are not power algebras.
Therefore we have
to introduce the notion of powers and show how to calculate
with these powers. This point will be studied in the next sections.

\section{Powers and Tree Graphs, Forms and Power Series}
\label{PowTre}
If $\cB $ is not a power algebra the power $x^n$ is not uniquely defined
for $x\in \cB $ and $n\in \bN $. In this section we will generalize
the notion of powers for general non associative algebras
(cf.~\cite{B66}.)

We will see that the different ways of multiplying $n$ copies
of an element $x$ correspond to binary trees. Trees do also occur in
renormalization theory of field theoretic models. These tree are
related to a non-associative algebra structure.

Let $\cF $ be a set provided with an addition + and
containing a zero element such
that $\cF $ is generated by a unit element $1\ne 0$. The addition is neither
assumed to be commutative nor associative.
$\cF $ is called {\em the set of all
non associative integers}. The elements of $\cF $ consist of
sums of the unit element $1$, e.~g.,
\be
(1+1)+1,\  (1+1)+(1+1),\  (1+(1+1))+1.
\ee
Let $\cP (\cB )$ be the set of all products of elements of $\cB $.
Define a homomorphism
\be
h:\, \cP (\cB ) \rightarrow \cF ,
\ee
where in $h(P)$ is defined by replacing in the product $P$ each factor
by 1 and each multiplication $\times $ by addition +. We define
$h(\eins ) := 0$. $h(P)$ is called the {\em form of the product $P$}
(cf.~\cite{B66}).
We may identify each element $S\in \cF $ containing $n$ unit elements 1
by a rooted tree with $n$ final vertices such that all vertices which
are not final have 2 successors. Denote the set of all such defined
(binary) trees by
$T_n^{(2)}$.

\begin{lemma} \label{numtrees}
For $n\ge 1$, the number of all binary trees with $n$ final vertices is
\be
|T_n^{(2)}| =\frac{(2n-2)!}{(n-1)!\, n!}.
\ee
\end{lemma}
\vskip5mm\noindent
\underline{\sl Proof:\,} We have the following recursive relations
\be
|T_1^{(2)}| =1, \qquad |T_n^{(2)}| = \sum_{k=1}^{n-1} |T_{n-k}^{(2)}|\,
    |T_k^{(2)}|,\ n\ge 2.
\ee
The generating function $F$ of the number of binary trees $|T_n^{(2)}|$
\be
F(x) := \sum_{n:\, n\ge 1} |T_n^{(2)}|\, x^n
\ee
obeys
\be
F^2(x) = F(x) - x.
\ee
This implies, supposing $|4x|<1$,
\be \label{genpow}
F(x) = \half (1- \sqrt{1-4x}) = -\half \sum_{n:\, n\ge 1} {\half \choose n}
           (-4)^n x^n.
\ee
Since
\be
{\half \choose n} = \frac{\half (-\half ) \cdots (\half -n+1)}{n!} =
        \frac{(-1)^{n-1}}{2^n\, n!} 1\cdot 3\cdot 5 \cdots (2n-3),
\ee
eq.~(\ref{genpow}) implies
\be
|T_n^{(2)}| = \frac{2^{n-1}}{n!} 1\cdot 3\cdot 5 \cdots (2n-3) =
                \frac{(2n-2)!}{n!\, (n-1)!}. \  \  \Box
\ee
For a form $S\in \cF $ the {\em degree $\delta (S)$} of $S$ is the
number of unit elements in the sum $S$. Denote
\be
\cF_n := \{ S\in \cF |\, \delta (S) =n\}
\ee
By Lemma \ref{numtrees} we have $|\cF_n| = \frac{(2n-2)!}{n!\, (n-1)!}$.

Let us define an equivalence class in $B$ by
\be
a \sim b \Leftrightarrow h(a)=h(b),
\ee
for all $a,b\in \cB $. Consider $x\in \cB $. Each equivalence class can
be represented by a form $S\in \cF $ where each factor in the product is
equal to $x$. Denote this product by $x^S.$ Define for two forms $S_1, S_2
\in \cF $ the power $x^{S_1 S_2} = (x^{S_1})^{S_2}$ by replacing $x$ by
$x^{S_1}$ in the product $x^{S_2}$. This product is associative but not
commutative
\be
(S_1S_2)S_3 = S_1(S_2S_3), \qquad S_1S_2 = S_2S_1.
\ee
For a symbol $X$ consider formal power series
\be
F(X) := \sum_{S:\, S\in \cF } f(S)\, X^S,\  \  f(S)\in \bC .
\ee
Denote the set of all such formal power series by $\bC_{\cF }[[X]]$.
For $F,G\in \bC_{\cF }[[X]]$ define addition and multiplication by
\bea
(F+G)(X) &:=& \sum_{S:\, S\in \cF } (f(S)+g(S))\, X^S
\nonu\\
(FG)(X) &:=& \sum_{T:\, T\in \cF } \left (\sum_{S_1,S_2 \in \cF :
              \atop T=S_1+S_2} f(S_1) g(S_2) \right ) \, X^T.
\eea
Replacing $X$ in $F(X)$ by an element $x\in \cB $ we get a formal power
series denoted by $F(x)$. We have
\be
(FG)(x) = F(x) \times G(x).
\ee
In the next section we will see how to express quasi-inverses
and quasi-roots in terms of power series.\section{Quasi-Roots and
Quasi-Inverses}
\label{AnaNon}
In algebras without unit elements the definition of an inverse
element is not possible. In this case the notion of an inverse can
be replaced by the notion of a quasi-inverse.

Let $\cB $ be an algebra without unit element. We call an element $x$ a
{\em quasi-inverse} of $y$ iff
\be
x\circ y := x+y+x\times y =0.
\ee
The multiplication $\circ $ is called {\em quasi-product}.
This definition can be motivated by the fact that if $x$ is a quasi-inverse
of $y$ then $\eins +x$ is the inverse of $\eins +y$ in $\cB_{\eins}$, where
$\cB_{\eins}$ is the algebra $\cB $ adjuncted with a unit element $\eins $
defined by eq.(\ref{einsad}).

We call $y\in \cB $ a {\em quasi-root} of $x\in \cB $ iff
\be
y\circ y = 2y + y\times y =x.
\ee
This definition can be motivated by $(\eins +y)\times (\eins +y)= \eins
+x$.
\begin{lemma} \label{alinv}
Let $\cB $ be a commutative algebra and
$y\in \cB $ be the quasi-inverse of $x\in \cB $. Then we
have the following formal power series expansion
\be \label{quadsol}
y = \sum_{n=1}^\infty (-1)^n\, x^{1_n},
\ee
where the forms are recursively defined by
\be \label{formrec}
1_1:=1 ,\qquad 1_{n+1} := 1_n+1,
\ee
for all $n\ge 1$.
Furthermore, suppose that there exists a norm $\Vert \cdot
\Vert $ in $\cB $ and $q\in \bR_+$ such that
\be
\Vert a\times b\Vert \le q \Vert a\Vert \, \Vert b \Vert
\ee
for all $a,b\in \cB $. Suppose that $q\, \Vert x\Vert <1$.
Then, we have
\be
\Vert y\Vert \le \frac{\Vert x\Vert }{1-q\, \Vert x\Vert }.
\ee
\end{lemma}
We get an analogous lemma for the quasi-root.
\begin{lemma} \label{alsqrt}
Let $\cB $ be an algebra and
$y\in \cB $ be the quasi-root of $x\in \cB $.
Then we have the following formal power series expansion
\be \label{quadsol2}
y = \sum_{S:\, S\in \cF -\{ 0\} } u(S)\, x^S,
\ee
where the coefficients $u(S)$ are recursively determined by
\be \label{usrec}
u(1) = \half ,\qquad u(T) = -\half \sum_{S_1,S_2 \in \cF -\{0\} :
              \atop T=S_1+S_2} u(S_1) u(S_2)
\ee
for $T\in \cF -\{0,1\} .$
Furthermore,
\be
u_n := \sum_{T:\, T\in \cF_n} |u(T)| \le \frac{1\cdot 3\cdot 5 \cdot
         \ldots \cdot (2n-3)}{2^n\, n!} = 2^{-2n+1}\,
          \frac{(2n-2)!}{n!\, (n-1)!} .
\ee
We have $u_n \le \frac{1}{2(n-1)}$ for $n\ge 2$.
\end{lemma}
\vskip5mm\noindent
\underline{\sl Proof:\,}
Let $y$ be the quasi-root of $x$. Then we have
\be
(\eins +y)\times (\eins +y) = \eins +x.
\ee
For $c\in \bR $, we have
\be
(1+\sum_{m:\, m \ge 1} q_m c^m )^2 =1+c, \qquad q_m := {\half \choose m} =
  \frac{(-1)^{n-1}}{2^n n!} 1\cdot 3\cdot 5 \cdot \ldots \cdot (2n-3).
\ee
Thus
\be
q_1 =\half ,\qquad q_n = -\half \sum_{m=1}^{n-1} q_{n-m} q_m .
\ee
We use induction in $n.$ Suppose that $u_m \le (-1)^{m-1} q_m$ for all $m<n$.
By eq.~(\ref{usrec}), we have
\bea
u_n &=& \sum_{T:\, T\in \cF_n} |u(T)| \le
        \half \sum_{T:\, T\in \cF_n} |u(T)| \sum_{S_1,S_2 \in \cF -\{0\} :
              \atop T=S_1+S_2} |u(S_1)|\, |u(S_2)|
\nonu\\ &\le &
\half \sum_{m=1}^{n-1} (\sum_{S_1:\, S_1\in \cF_n} |u(S_1)|)
       (\sum_{S_2:\, S_2\in \cF_n} |u(S_2)|) =
\half \sum_{m=1}^{n-1} u_{n-m}\, u_m .
\eea
By induction hypothesis we have
\be
u_n \le -\frac{(-1)^{n-1}}{2} \sum_{m=1}^{n-1} q_{n-m} q_m =
   (-1)^{n-1} q_n.
\ee
This proves the assertion. $\  \   \Box $

\begin{lemma} Consider $F(X) = \sum_{S:\, S\in \cF -\{ 0\} } f(S)\, X^S
\in \bC_{\cF }[[X]]$ and suppose that there exists a norm $\Vert \cdot
\Vert $ in $\cB $ and $q\in \bR_+$ such that
\be
\Vert a\times b\Vert \le q \Vert a\Vert \, \Vert b \Vert
\ee
for all $a,b\in \cB $. Then, we have for $x\in \cB $
\be \label{Fxbou}
\Vert F(x)\Vert \le \frac{1}{q} \, \left [ \sum_{n:\, n\ge 1} f_n\,
     (q\Vert x\Vert )^n \right ],
\ee
where
\be \label{deffn}
f_n := \sum_{S:\, S\in \cF_n} |f(S)|.
\ee
Let $y\in \cB $ be the quasi-inverse of $x\in \cB $.
Suppose that $q\Vert x\Vert <1$.
Then, we have
\be \label{vyvbou}
\Vert y\Vert \le \frac{\Vert x\Vert }{2}\left ( 1 + \ln (1-q\Vert x\Vert )
  \right ).
\ee
\end{lemma}
\vskip5mm\noindent
\underline{\sl Proof:\,} We have
\be
\Vert F(x)\Vert \le \sum_{n:\, n\ge 1} \sum_{S:\, S\in \cF_n} |f(S)|
        q^{n-1} \Vert x\Vert^n .
\ee
This and definition (\ref{deffn}) proves eq.~(\ref{Fxbou}).
By Lemma \ref{alsqrt} and the bound eq.~(\ref{Fxbou}) we see
\be
\Vert y\Vert \le \frac{1}{q} \sum_{n:\, n\ge 1} u_n (q\Vert x\Vert )^n
\le \frac{1}{2q}\left ( q\Vert x\Vert + \sum_{n:\, n\ge 2}
   \frac{(q\Vert x\Vert )^n}{n-1} \right ).
\ee
This proves eq.~(\ref{vyvbou}). $\  \  \Box $
\chapter{Introduction to Renormalization Group Algebras}

\section{Examples of Hierarchical Renormalization Group Algebras}
\label{DefHie}
In this section we study examples of hierarchical renormalization group
algebras. We will show how these examples can be related to hierarchical
renormalization group transformations.
Fixed points of the renormalization group transformation (RGT)
 are the idempotents of the renormalization group algebra.

In the following we consider the space $\cB := L_2(e^{-\frac{\Phi^2}{2\g }}\,
d\Phi )$.
For $L>1$, $\beta $, $\g >0$ and
the (not normalized) {\em hierarchical
renormalization group transformation} $\cR_{\g ,L^d}^\beta $ in
$d$ dimensions is defined by
\be \label{rgtrans}
\cR_{\g ,L^d}^\beta (F)(\Psi ) = \iGaug F(\Phi + \beta \Psi )^{L^d},
\ee
where $\Gaug $ is a Gaussian measure with mean zero defined by
\be
\Gaug := (2\pi \g )^{-1/2} d\Phi \, \exp \{ -\frac{1}{2\g } \Phi^2\} .
\ee
The RGT depends on the three parameters $\beta ,\  L^d$ and $\g .$
$\beta $ is called the {\sl scaling parameter}, $L^d$ the
{\sl volume factor}, and $\g $ the {\sl (free) covariance}
of the RG transformation.
The volume factor and the scaling parameter
obey the relation
\be \label{betaeq}
\beta = L^{1-\frac{d}{2}}.
\ee
Define a mapping $E_\g :\, \cB
\rightarrow \cB $ by $E_\g := \cR_{\g ,1}^1$, i.~e.
\be
E_\g (F)(\Phi ) := \iGau{\g }{\zeta } F(\zeta +\Phi )
\ee
and the Gaussian expectation value $<\ >_\g :\, \cB  \rightarrow \bR $
by
\be
<F>_\g := E_\g (F)(0) = \iGau{\g }{\zeta } F(\zeta ).
\ee
We are interested in the existence and construction of fixed points
$F^*$ of the RG transformation (\ref{rgtrans})
\be
\cR_{\g ,L^d}^\beta (F^*) = F^*.
\ee
Three fixed points are immediately found. The zero fixed point
$F_0:=0$, the ultraviolet fixed point $F_{UV}:=1$ and the high
temperature fixed point $F_{HT}$ defined by
\be
F_{HT}(\Psi ) := L^{\frac{1}{L^{d}-1}} \exp \bigl \{ -\frac{L^2-1}{2\g L^d}
                                                 \Psi^2 \bigr \} .
\ee
The high temperature fixed point turns out to be stable. The most
important problem is that of finding non-trivial unstable fixed
points. These will be called infrared fixed points. They are related
to phase transitions in our hierarchical model in the infrared limit.

Thus we are looking for some other nontrivial fixed points $F_{IR}$.
That $F_{HT}$ is a fixed point can be shown as an application of the
following Lemma
\begin{lemma}\label{lemma1.1}
For all $c\ne -\frac{1}{2\g }$ and
$F\colon \bR \rightarrow \bR $, such that the following Gaussian
integrals exist, we have
\be
E_\g (e^{-c(\cdot )^2} F(\cdot ))(\Psi ) =
\cL_c^{1/2} \exp\{ -c\cL_c \Psi^2 \} E_{\cL_c \g }
(F)(\cL \Psi ),
\ee
where $\cL_c := (1+2\g c)^{-1}$.
\end{lemma}
\vskip5mm\noindent
The foregoing Lemma \ref{lemma1.1} implies the following Corollary.
\begin{corollary} \label{quadextr} Define
\be
\cU_c^\cN (F)(\Phi ) := \cN \, \exp \{ c\Phi^2 \} \, F(\Phi ).
\ee
For all $c\ne \frac{1}{2\g L^d}$ we have
\be
\cU_c^\cN \circ \cR_{\g ,L^d}^\beta \circ (\cU_c^\cN )^{-1} =
\cU_{c(1-L^d \cL \beta^2)}^{\cL^{\frac{L^d}{2}}\cN^{-L^d+1}} \circ
\cR_{\cL \g ,L^d}^{\beta \cL } ,
\ee
where
\be
\cL := \frac{1}{1+2\g L^d c}.
\ee
\end{corollary}
\vskip5mm\noindent
\underline{\sl Proof:\,} The assertion is proven by
\bea
\cR_{\g ,L^d}^\beta \circ (\cU_c^\cN )^{-1} (F)(\Psi ) &=&
  \iGaug [\cN^{-1} \exp \{ -c(\Phi +\beta \Psi )^2 \} F(\Phi +\beta \Psi )
   ]^{L^d}
  \nonu\\ &=&
 \cN^{-L^d}\, \iGaug \exp \{ -cL^d(\Phi +\beta \Psi )^2 \}
      F^{L^d}(\Phi +\beta \Psi )
 \nonu\\ &=&
\cL^{\frac{L^d}{2}} \cN^{-L^d}
   \exp \{ -cL^d\cL \beta^2 \Psi^2 \} \iGau{\cL \g }{\Phi }
  F^{L^d}(\Phi +\beta \cL \Psi )
   \nonu\\ &=&
\cU_{-cL^d \cL \beta^2)}^{\cL^{\frac{L^d}{2}}\cN^{-L^d}} \circ
\cR_{\cL \g ,L^d}^{\beta \cL }(F)(\Psi ) . \  \  \Box
\eea
We introduce a similarity transformation which preserves the form
of the renormalization group transformation eq.~(\ref{rgtrans})
but changes the parameters in a favorable manner.

For a transformation
$\cS :\, \cB \rightarrow \cB $, define
the equivalent renormalization group transformation
\be
\cT_\cS [\cR ] := \cS \circ \cR \circ \cS^{-1}.
\ee
The following special similarity transformation were used by Koch and Wittwer
\cite{KW86,KW91} for an analysis of the double-well fixed point in 3 dimensions
by means of a rigorous beta-function technique.
\be
\cS_{HT}:\, \cB \rightarrow \cB , \qquad
\cS_{HT} (F) (\Phi ) := \frac{F(\Phi )}{F_{HT}(\Phi )} = Z(\Phi ).
\ee
Then Corollary \ref{quadextr} implies
\be
\cT_{\cS_{HT}}[\cR_{\g ,L^d}^\beta ] = \cR_{\g^{'} ,L^d}^{\beta^{'}},
\ee
where
\be
\beta^{'}:= L^{-1-\frac{d}{2}},\qquad \g^{'}:= L^{-2}\g ,
\ee
if $\beta $ is defined by eq.\ (\ref{betaeq}).
The result of the transformation $\cT_{\cS_{HT}}$ is that the scaling
parameter $\beta $ becomes smaller. Furthermore, the logarithm of fixed
points becomes more convex which leads to a simpler treatment of the
large field problem.

In the following, we will study the transformed RG transformation
$\cR_{\g^{'},L^d}^{\beta^{'}}$ for the case
$L^d=2$.

Let us remark that the restriction to $L^d=2$ is not necessary
for the use of methods
in this paper. We will use only the condition $L^d=2$ in order to keep
the notations simpler. All results can be easily generalized to the
case $L^d\in \{ 2,3,\ldots \} .$

We will write $\cR_{\g^{'}}^{\beta^{'}} =
\cR_{\g^{'},1}^{\beta^{'}}$. Obviously, $\cR_{\g^{'},2}^{\beta^{'}}(F) =
\cR_{\g^{'}}^{\beta^{'}} (F^2).$
The fixed points of the original RGT are in one to one correspondence
with the fixed points of the new RGT.
The infrared fixed points $F_{IR}$
of $\cR_\g^\beta $ and $Z_{IR}$ are related by
\be
F_{IR}(\Phi ) = F_{HT}(\Phi ) \, Z_{IR}(\Phi ) .
\ee
The RG transformation $\cR_{\g^{'},L^d}^{\beta^{'}}$,
$\beta^{'}:= L^{-1-\frac{d}{2}}$, has the high temperature fixed point
$Z_{HT}=1$ and the UV fixed point $Z_{UV}$, where
\be
Z_{UV}(\Psi ) := L^{-\frac{1}{L^d-1}}\, \exp \{ \frac{1-L^{-2}}{2\g^{'}L^d}
\Psi^2 \} =
  2^{-\frac{1}{d}}\, \exp \{ \frac{1-2\beta^{'2}}{4\g^{'}} \Psi^2 \} .
\ee
The high temperature fixed point $F_{HT}$ of $\cR_\g^\beta $ reads
\be
F_{HT}(\Phi ) = 2^{\frac{1}{d}}\, \exp \{ -\frac{2\beta^{2}-1}{4\g }
\Phi^2 \}
\ee
and we have
\be
\cR_\g^\beta \left ((F_{HT}\cdot A)\cdot (F_{HT}\cdot B)\right ) =
  F_{HT}\, \cR_{\g '}^{\beta '}(A\cdot B).
\ee

In the following let us define some multiplication on $\cB $.
For $A,B \in \cB $ define a $\cdot $-product by
\be \label{cdotdef}
A\cdot B (\Phi ):= A(\Phi ) \, B(\Phi ).
\ee
$(\cB ,\cdot )$ defines a commutative and associative algebra.

Let $\cS_\delta :\, \cB \rightarrow \cB $
be the scaling transformation
\be
\cS_\delta (F)(\Phi ) = F(\delta \Phi ).
\ee
For $A,B \in \cB $ define a $\ast $-product by
\be \label{astdef}
A\ast_{\beta ,\g} B := \cS_\beta E_\g (A\cdot B)= \cR_{\g }^\beta (A\cdot B).
\ee
$(\cB ,\ast )$ defines a commutative algebra.
We will omit the second index $\g $ in $\ast_{\beta ,\g} $ if no confusion
can arise.

Let us introduce Wick-ordered functions
by
\be
:F(\Phi ):_\g := \exp \{ -\frac{\g }{2} \frac{\partial^2}{\partial
  \Phi^2} \} F(\Phi ) .
\ee
Note that $E_{-\g }(F) = :F:_\g $ and
\be \label{convol}
E_{\g_1+\g_2} = E_{\g_1} \circ E_{\g_2},
\ee
where $\circ $ denotes the composition of two functions.

The Wick-ordered monomials are expressed by a sum of monomials in the
following Lemma.

\begin{lemma}\label{wickmon} For $n\in \bN $ we have
\be
:\Phi^n:_\g = \sum_{k:\, n-k\in 2\bN} (-\frac{\g }{2})^{\frac{n-k}{2}}
   {n \choose k}
               \frac{(n-k)!}{(\frac{n-k}{2})!}\, \Phi^k.
\ee
\end{lemma}
\vskip5mm\noindent
\underline{\sl Proof:\,} We have
\bea
:\Phi^n:_\g &=& \exp \{ -\frac{\g }{2} \frac{\partial^2}{\partial
               \Phi^2} \} \Phi^n \nonumber\\ &=&
\sum_{m=0}^{[\frac{n}{2}]} (-\frac{\g }{2})^m \frac{n(n-1)\cdots
       (n-2m+1)}{m!} \Phi^{n-2m} \nonumber\\ &=&
\sum_{k:\, n-k\in 2\bN}  (-\frac{\g }{2})^{\frac{n-k}{2}}
{n \choose k} \frac{(n-k)!}{(\frac{n-k}{2})!} \Phi^{k}.
\eea
This implies the assertion. $\  \  \Box $

\begin{lemma}\label{wickquad} For $c\ne -\frac{1}{2\g }$ we have
\be
:\exp \{ c\Phi^2 \} \, G(\Phi ):_\g =
\cL_c^{\half} \exp \{ c\cL_c \Phi^2 \} :G:_{\cL_c \g }(\cL_c \Phi ),
\ee
where
\be
\cL_q := (1+2\g q)^{-1}.
\ee
\end{lemma}
\vskip5mm\noindent
\begin{lemma}\label{wickortho} For all $m,n\in \bN $ we have the
orthogonality relation
\be \label{ortho}
<:\Phi^m:_\g \, :\Phi^n:_\g >_\g = \delta_{m,n} \, \g^m m!.
\ee
Furthermore, for all $a_i \in \bC$, $i\in \{ 1,\ldots ,n\} $, we have
\be \label{prodbol}
<\prod_{i=1}^n :\exp \{ a_i\Phi^2 \} :_\g >_\g = \prod_{i,j:\, i<j}
  \exp \{ \g a_i a_j \}
\ee
and
\be \label{prodquad}
<:\exp \{ a\Phi^2 \} :_\g \cdot :\exp \{ b\Phi^2 \} :_\g >_\g =
  (1-4\g^2 ab )^{\half}.
\ee
\end{lemma}
\underline{\sl Proof:\,} First of all, we will prove eq.~(\ref{prodbol}).
By the definition of Wick-ordered products, we obtain
\be \label{1.44}
< \prod_{i=1}^n : \exp \{ a_i\Phi \} :_\g >_\g = \biggl ( \prod_{i=1}^n
\exp \{-\frac{\g }{2} a_i^2\} \biggr )\, <\exp \{
  (\sum_{j=1}^n a_j)\Phi \} >_\g .
\ee
Since, for all $a\in \bC $,
\be
<\exp \{ a\Phi \} >_\g = \exp \{ \frac{\g }{2} a^2\} ,
\ee
eq.~(\ref{1.44}) implies
\be
< \prod_{i=1}^n : \exp \{ a_i\Phi \} :_\g >_\g = \exp \{ -\frac{\g }{2}
[\sum_{i=1}^n a_i^2 -(\sum_{j=1}^n a_j )^2 ]\} .
\ee
This proves eq.~(\ref{prodbol}). We will prove eq.~(\ref{ortho}).
Using eq.~(\ref{prodbol}) we obtain
\bea
<:\Phi^m:_\g \, :\Phi^n:_\g >_\g &=&
\frac{\partial^m}{\partial a^m} \frac{\partial^n}{\partial b^n}
< :\exp \{ a\Phi \}:_\g \, :\exp \{ b\Phi \} :_\g >_\g \vert_{a=b=0}
\nonumber\\ &=&
\frac{\partial^m}{\partial a^m} \frac{\partial^n}{\partial b^n}
\exp \{ \g ab\} \vert_{a=b=0} =  \delta_{m,n} \g^m m!.
\eea
We will prove eq.~(\ref{prodquad}). Using Lemma \ref{wickquad}, we obtain
\bea
\lefteqn{<:\exp \{ a\Phi^2 \} :_\g \, :\exp \{ b\Phi^2 \} :_\g >_\g =}
\nonumber\\ & &
=(\cL_a \cL_b)^{\half} \iGaug \exp \{ (a\cL_a +b\cL_b )\Phi^2 \}
\nonumber\\ & &
=(\cL_a \cL_b \cL_{-(a\cL_a +b\cL_b)})^{\half}
\nonumber\\ & &
=[(1+2\g a)(1+2\g b) (1 - \frac{2\g a}{1+2\g a} - \frac{2\g b}{1+2\g b}
  )]^{-\half }
\nonumber\\ &
=(1-4\g^2 ab )^{\half}. \  \  \Box
\eea
\vskip5mm\noindent
We have the following scaling relation for Wick ordered functions
\be \label{wickscale}
:F:_{\delta^2 \g} (\delta \Phi ) = :\cS_\delta (F):_\g (\Phi ).
\ee
Eq.~(\ref{convol}) implies
\be
:E_\g (F):_\g = E_\g (:F:_\g ) = F .
\ee
Thus $E_\g (\cdot )$ is the inverse of Wick-ordering $:(\cdot ):_\g .$
For two functions $A,B \in \cB $ define a product
\be \label{prodef}
A \times_\beta B := \cS_\beta E_\g (:A:_\g \cdot :B:_\g ).
\ee

\noindent
Let us remark that eq.~(\ref{prodef}) is equivalent to
\be \label{wickprod}
:\cS_\beta^{-1} (A \times_\beta B):_\g  = :A:_\g \cdot :B:_\g.
\ee

\begin{lemma}\label{rgtalg} For $A,B\in \cB $, we have
\be \label{prodrg}
:A\times_{\beta ,\g }B:_\g = R_{\g (1-\beta^2)}^\beta (:A:_\g \cdot :B:_\g ).
\ee
\end{lemma}
\vskip5mm\noindent
\underline{\sl Proof:\,} We have,
\be
R_{\g (1-\beta^2)}^\beta (:A:_\g \cdot :B:_\g ) =
R_{\g (1-\beta^2)}^\beta (:A\times_1 B:_\g ).
\ee
Therefore,
\be
R_{\g (1-\beta^2)}^\beta (:A:_\g \cdot :B:_\g ) (\Phi )=
:A\times_1 B:_{\beta^2 \g }(\beta \Phi ).
\ee
This implies, using relation (\ref{wickscale}), the assertion
eq.~(\ref{prodrg}). $\   \   \Box $

\abstand
$(\cB ,\times )$ defines a commutative algebra. This algebra is associative
for the case $\beta =1$ and nonassociative for the case $\beta \ne 1$.

The following Lemma shows a relation of the $\times $-and
$\ast $-product.

\begin{lemma} \label{relasttimes}
For $\beta \ne 1$, we have
\be
: A \times_{\beta ,\g }B:_\g = :A:_\g \ast_{\beta ,\g (1-\beta^2)}
                           :B:_\g .
\ee
\end{lemma}
\vskip5mm\noindent
\underline{\sl Proof:\,} We have, using Lemma \ref{rgtalg}
\be
:A\times_{\beta ,\g }B:_\g = \cR_{\g (1-\beta^2)}^\beta (:A:_\g
        \cdot :B:_\g ).
\ee
The definition eq.~(\ref{astdef}) implies the assertion. $\  \  \Box $

The UV fixed point for
$\cR_{\g (1-\beta^2),2}^{2^{-\frac{2+d}{2d}}}$ reads
\be
Z_{UV}(\Phi ) = 2^{-\frac{1}{d}} \exp \{ c_* \Phi^2 \} ,
\ee
where
\be
c_* := \frac{1-2\beta^2}{4\g (1-\beta^2)}, \qquad \beta :=
2^{-\frac{2+d}{2d}}.
\ee

\noindent
Consider now the case $\beta =1$. The mapping
$E_{-\g }:\, (\cB ,\times ) \rightarrow (\cB ,\cdot )$ is
an algebra-isomorphism, i.~e.~,
\be
E_{-\g }(A\times B) = E_{-\g }(A) \cdot E_{-\g }(B),
\ee
for all $A,B\in \cB $. This algebra-isomorphism shows that $(\cB ,
\times_\beta )$ is associative for the case $\beta =1$.
The function $\eins \in \cB $ defined by
\be
\eins (\Phi ):= 1
\ee
is the unit element of $(\cB ,\times )$ resp.~$(\cB ,\cdot )$
For $\beta \ne 1$ there is no unit element of $(\cB ,\times )$.
We may adjoin to $(\cB ,\times )$ a unit element $\eins $ by defining
the multiplication of two elements $a\eins +A$, $b\eins +B$, for $a,b \in
\bC $ and $A,B\in \cB $ by
\be
(a\eins + A)\times (b\eins + B):= (ab)\eins + (aB+bA+A\times B).
\ee
This defines the algebra $(\cB \oplus \{ \eins \} ,\times )$ with unit
element $\eins $.

\section{Constants of Multiplication}
\label{ConMul}
In this section we present the constants of multiplication for the algebras
defined in section 3.~1. Useful estimations and relations are derived
for the constants of multiplication. The form of the constants
of multiplication depends on the basis chosen. We will consider different
bases. The first basis is given by simple powers of $\Phi .$ Later we
will also consider normal ordered products.

Consider for the algebra $\cB $
a basis $\{ f_m |\, m\in \bN \} $ defined by
\be
f_m  := \frac{\Phi^m}{\g^{\frac{m}{2}}}.
\ee
The {\sl constants of multiplication} $M_l^{mn}$ for the
$\cdot $-multiplication are given by
\be \label{algmul1}
f_m \cdot f_n =
\sum_{l:\, l\in \bN } M_l^{mn} \, f_l
\ee
where
\be
M_l^{mn} = \delta_{m+n,l}.
\ee
The {\sl constants of multiplication} $S_l^{mn}$ for the
$\ast $-multiplication are defined by
\be \label{algmul}
f_m \ast_\beta f_n = \beta^l\,
\sum_{l:\, l\in \bN } S_l^{mn} \, f_l.
\ee
\begin{lemma}\label{strucoe} For a triple $(l,m,n) \in \bN^3$
such that $m+n-l\in 2\bN $ , we have
\be
S_l^{mn} = 2^{\frac{l-m-n}{2}}
                   \frac{(m+n)!}{[\half (m+n-l)]!\, l!}
\ee
and $S_l^{mn} = 0$ if $m+n-l\in 2\bN +1$ or $l>m+n.$
\end{lemma}
\vskip5mm\noindent
\underline{\sl Proof:\,} We have,
\be
<(\Phi +\beta \Psi )^{m+n}>_\g = \sum_l \beta^l {m+n \choose l}
<\Phi^{m+n-l} \Psi^l>_\g .
\ee
Since
\be
<\Phi^{m+n-l}>_\g = (\frac{\g}{2})^{\frac{m+n-l}{2}}
                   \frac{(m+n-l)!}{[\half (m+n-l)]!},
\ee
for $m+n-l \in 2\bN $, and $=0$ otherwise, the assertion follows.
$\  \  \Box $

\abstand
We consider the subalgebra $\cB^{(e)} \subset \cB $ of even functions.
A basis $\{ v_m|\, m\in \bN \} $ is defined by $v_m :=f_{2m}$. For this
subalgebra of $\cB $ the constants of multiplication are
\be
\cS_l^{mn}:= S_{2l}^{2m\,2n}.
\ee
For the following Lemma define
\be
A_l^{mn} := \sqrt{\frac{(2l)!}{(2m)! (2n)!}} \, \cS_l^{mn}
\ee
\begin{lemma} \label{cbouast} $A_l^{mn}$ obeys the bound
\be
A_l^{mn} \le \sqrt{2(m+n) \choose 2m} {m+n \choose l}.
\ee
\end{lemma}
\vskip5mm\noindent
\underline{\sl Proof:\,} We obtain
\be \label{Aequ}
A_l^{mn}  = 2^{-2(m+n-l)} {m+n \choose l}^2 \, {2(m+n) \choose 2m} \,
     \frac{[2(m+n)]!}{(m+n)!^2}\, \frac{l!^2}{(2l)!} .
\ee
For $k\in \bN $ we have
\be
\frac{1}{4} \, \frac{k!^2}{(2k)!} \le \frac{(k+1)^2}{(2k+1)(2k+2)}
           \, \frac{k!^2}{(2k)!} =  \frac{(k+1)!^2}{[2(k+1)]!} .
\ee
Therefore, since $l\le m+n$,
\be \label{lfacest}
\frac{l!^2}{(2l)!} \le 4^{m+n-l} \frac{(m+n)!^2}{[2(m+n)]!}.
\ee
Insertion of the estimation eq.~(\ref{lfacest}) on the rhs of
eq.~(\ref{Aequ}) proves the assertion. $\  \   \Box $

\begin{lemma} For all $a,b\in \bC $, we have
\be
\exp \{ a\Phi \} \ast_\beta \exp \{ b\Phi \} =
  \exp \{ \frac{\g }{2} (a+b)^2 \} \exp \{ \beta (a+b)\Phi \}
\ee
and
\be
\sum_{m,n:\, l\le m+n} S_l^{mn} \frac{a^mb^n}{m!n!} = \frac{(a+b)^l}{l!}
  \, \exp \{ \frac{(a+b)^2}{2} \} .
\ee
Suppose that $a,b\in \bR $, $2(a+b)<1$.
Then we have
\be
\exp \{ a\Phi^2 \} \times_\beta \exp \{ b\Phi^2 \} =
  \frac{1}{\sqrt{1-2(a+b)}} \, \exp \{ \frac{a+b}{1-2(a+b)}\beta^2 \Phi^2 \}
\ee
and
\be
\sum_{m,n:\, l\le m+n} \cS_l^{mn} \frac{a^mb^n}{m!n!} =
   \frac{1}{\sqrt{1-2(a+b)}} \, \frac{[\frac{a+b}{1-2(a+b)}]^l}{l!}.
\ee
\end{lemma}
Consider for the algebra $(\cB ,\times )$
a basis $\{ e_m |\, m\in \bN \} $ defined by
\be
e_m  := \frac{:\Phi^m:_\g}{\g^{\frac{m}{2}}} .
\ee
Using Lemma \ref{wickmon} we obtain
\begin{lemma} For $n,k\in \bN $ and $\alpha \in \bR $ define
\be
E_{nk}(\alpha ) := \left\{ \begin{array}{r@{\quad:\quad}l}
       (\frac{\alpha }{2})^{\frac{n-k}{2}} {n\choose k}
       \frac{(n-k)!}{(\frac{n-k}{2})!}
         & n-k\in 2\bN \\ 0 & n-k \not\in 2\bN
         \end{array} \right.
\ee
Then we have
\bea
e_n &=& \sum_{k:\, n-k \in 2\bN } E_{nk}(-1) f_k,
\nonu\\
f_n &=& \sum_{k:\, n-k \in 2\bN } E_{nk}(1) e_k .
\eea
\end{lemma}

\noindent
The {\sl constants of multiplication} $C_l^{mn}$ are defined by
\be \label{algmulwick}
e_m \times_\beta e_n = \beta^l\,
\sum_{l:\, l\in \bN } C_l^{mn} \, e_l.
\ee
For a triple $(l,m,n) \in \bN^3$ define the ``dual'' triple
$(l^{'}, m^{'}, n^{'})$ by
\be \label{dual}
l^{'} := m+n-l,\qquad  m^{'} := l+n-m,\qquad n^{'} := m+l-n.
\ee
\begin{lemma}\label{constmul} For a triple $(l,m,n) \in \bN^3$
such that the ``dual'' triple $(l^{'}, m^{'}, n^{'})\in 2\bN^3 $
has even entries, we have
\be
C_l^{mn} = \frac{m! n!}{(\half l^{'})!\, (\half m^{'})!\,
                      (\half n^{'})!}
\ee
and $C_l^{mn} = 0$ if $(l^{'}, m^{'}, n^{'})\notin 2\bN^3 $
or $l<|m-n|$ or $l>m+n.$
Furthermore, we have
\be
\sum_{l:\, l\in \bN } C_l^{mn}\, C_k^{lq} =
\sum_{l:\, l\in \bN } C_l^{nq}\, C_k^{ml},
\ee
i.~e., the algebra $(\cB ,\times )$ is associative for $\beta =1$.
\end{lemma}
\vskip5mm\noindent
\underline{\sl Proof:\,} We have, using eq.~(\ref{algmulwick}),
\be
<:\Phi^l: \cdot :\Phi^m:\cdot :\Phi^n:>_\g =
\sum_{k \ge 0} C_k^{mn} \, < :\Phi^k: \cdot :\Phi^l : >_\g =
\sum_{k \ge 0} C_k^{mn} \, < :\Phi^k: \cdot :\Phi^l : >_\g .
\ee
The orthogonality relation of
Lemma \ref{wickortho}, eq.~(\ref{ortho}), yields
\be \label{ugl1}
<:\Phi^l:_\g \cdot :\Phi^m:_\g \cdot :\Phi^n:_\g >_\g =
C_l^{mn} \, \g^l \, l!.
\ee
Furthermore, Lemma \ref{wickortho}, eq.~(\ref{prodbol}), yields
\bea \label{ugl2}
<:\Phi^l:_\g \cdot :\Phi^m:_\g \cdot :\Phi^n:_\g >_\g &=&
\frac{\partial^{l+m+n}}{\partial a^l \partial b^m \partial c^n}
e^{\g (ab+bc+ac)} \vert_{a=b=c=0} \nonumber\\
&=& \sum_{m^{'} ,n^{'}  ,l^{'}  \in \bN }^{\prime }
\frac{\partial^{l+m+n}}{\partial a^l \partial b^m \partial c^n}
\g^{\frac{m+n+l}{2}} \frac{1}{m^{'} !n^{'}  ! l^{'}  !} \nonumber\\
& &(ab)^{n^{'}}  (bc)^{l^{'}} (ac)^{m^{'}} \vert_{a=b=c=0} ,
\eea
where the sum over $m^{'}  , n^{'}  , l^{'}   $, is restricted by
\be
l = m^{'}  +n^{'}  ,\qquad
m = l^{'}  +n^{'}  ,\qquad
n = l^{'}  +m^{'}  .
\ee
Eq.~(\ref{ugl1}) and eq.~(\ref{ugl2}) imply the assertion. $\  \  \Box $

We consider the subalgebra $\cB^{(e)} \subset \cB $ of even functions.
$\{ e_{2m}|\, m\in \bN \} $ is a basis of $\cB^{(e)}.$ For this
subalgebra of $\cB $ the constants of multiplication are
\be
\cC_l^{mn}:= C_{2l}^{2m\,2n}.
\ee

We will show that there exists a genetic basis for the algebra $(\cB^{(e)},
\times_1).$ By Lemma \ref{lweigen} this implies that there exists
a weight $\omega .$ Obviously, $\{ :\Phi^{2m}:_1,\  m\in \bN \} $ is
a basis. We have
\be
:\Phi^{2m}:_1 = 2^{-m}\, {\rm H}_{2m}(\frac{\Phi }{\sqrt{2}}),
\ee
where ${\rm H}_{n}$ is the $n$th Hermite-polynomial defined by
\be
\sum_{n=0}^\infty \frac{{\rm H}_n(\Phi )}{n!} \, z^n =
\exp \{ -z^2 +2z\Phi \} .
\ee
Thus,
\be
{\rm H}_{2m}(0) = (-1)^m \, \frac{(2m)!}{m!} .
\ee
Therefore,
\be
:\Phi^{2m}:_1\, :\Phi^{2n}:_1 = \sum_l \cC_l^{mn} \, :\Phi^{2l}:_1
\ee
implies, for $\Phi =0$,
\be
\frac{1}{(-2)^m}\, \frac{(2m)!}{m!}\, \frac{1}{(-2)^n}\, \frac{(2n)!}{n!}
=\sum_l \cC_l^{mn} \, \frac{1}{(-2)^l}\, \frac{(2l)!}{l!}.
\ee
Therefore, we see
\be
\sum_l \frac{\g (m) \, \g (n)}{\g (l)}\, \cC_l^{mn} =1,
\ee
for all $m,n\in \bN $, where
\be
\g (m) := \frac{(-2)^m\, m!}{(2m)!}.
\ee
This implies that $\{ \g (m)\, e_{2m},\  m\in \bN \} $ is a
genetic basis with constants of multiplication
$\frac{\g (m) \, \g (n)}{\g (l)}\, \cC_l^{mn}.$

\noindent
A weight $\omega :\, \cB \rightarrow \bC $ can be defined by
\be \label{defome}
\omega (x) = <\omega ,x> = \sum_m \omega_m\, x_m,
\ee
for all $x=\sum_m x_m\, e_{2m}$, where
\be
\omega_m := \g (m)^{-1}.
\ee
Suppose that the series on the rhs of eq.~(\ref{defome}) converges.
$\omega $ obeys the ``dual'' fixed point equation
\be
\sum_l \omega_l\, \cC_l^{mn} = \omega_m\, \omega_n .
\ee
Let us remark that the algebra $(\cB ,\times_\beta )$ is not weighted
if $\beta \ne 1.$ This can be shown as follows.
Suppose that there exists a weight $\omega .$
Consider
\be
\omega (e_0 \times_\beta e_{2m}) = \beta^{2m}\,
\omega (e_{2m}) = \omega (e_{0}) \cdot \omega (e_{2m})
\ee
for all $m\in \bN .$ This implies $\omega (e_{2m}) = 0$ for all $m\ge 1.$
Using
\be
\omega (e_{2m} \times_\beta e_{2n}) = \sum_l \cC_l^{mn} \, \omega (e_{2l})
= \cC_0^{mm} =0
\ee
we get a contradiction to $\cC_0^{mm}\ne 0.$ Thus, there exists no weight
$\omega $ for $(\cB ,\times_\beta )$, $\beta \ne 1.$

For $z\in \cB $ and the weight $\omega $ defined by eq.~(\ref{defome})
above we get the formal series
\be
\omega (z) = \sum_m \frac{(2m)!}{m!} (-\frac{1}{2})^m \, z_m ,\qquad
z = \sum_m z_m\, e_{2m} .
\ee
If $z$ is a fixed point of $(\cB ,\times_\beta )$, i.~e.~, $z\times_\beta z =
z,$ the following equation is valid
\be
\omega (\cS_{\beta^{-1}}(z)) =\omega^2(z).
\ee

For the following Lemma define
\be
B(l,m,n) := \sqrt{\frac{(2l)!}{(2m)! (2n)!}} \, \cC_l^{mn}.
\ee
We see that $B(l,m,n)$ is a symmetric function.

The following estimate is due to \cite{KW91}.
\begin{lemma} \label{cbou} For all $l,m,n\in \bN $ such that $|m-n| \le
l\le m+n$ we have
\be
B(l,m,n)\le \min_{(l^{'},m^{'},n^{'})\in \{ (l,m,n),(m,n,l),(n,l,m)\} }
\left (\sqrt{2l^{'} \choose l^{'}+m^{'}-n^{'}} {m^{'}+n^{'} \choose l^{'}}
\right ).
\ee
\end{lemma}
\vskip5mm\noindent
\underline{\sl Proof:\,} We obtain
\bea
B^2(l,m,n) &=&
   \frac{(2l)!}{(2m)!(2n)!} (\cC_{l}^{m n})^2
\nonu\\ &=&
 \frac{(2l)!}{(2m)!(2n)!}
\frac{(2m)!^2(2n)!^2}{(l+m-n)!^2(l+n-m)!^2}
\frac{1}{(m+n-l)!^2}.
\eea
This implies
\be
B^2(l,m,n) =
 {2l \choose l+m-n}
\frac{(2m)!(2n)!}{(l+m-n)!(l+n-m)!}
\frac{1}{(m+n-l)!^2}.
\ee
Thus
\be
B^2(l,m,n) =
 {2l \choose l+m-n} {m+n \choose l}^2
\frac{(2m)!(2n)!}{(m+n)!^2}
\frac{l!^2}{(l+m-n)!(l+n-m)!}.
\ee
We have to prove that
\be \label{hypoImnl}
I_{m,n,l}:= \frac{(2m)!(2n)!}{(m+n)!^2}
\frac{l!^2}{(l+m-n)!(l+n-m)!} \le 1
\ee
For $l \ge k$, define the function
\be
f_k(l):= \frac{l!^2}{(l-k)!(l+k)!}.
\ee
Since
\be
f_k(l) = \frac{(l+1-k)(l+1+k)}{(l+1)^2} f_k(l+1) =
\frac{(l+1)^2-k^2}{(l+1)^2} f_k(l+1) \le f_k(l+1),
\ee
and
\be
I_{m,n,l}= \frac{f_{m-n}(l)}{f_{m-n}(m+n)}, \qquad l \le m+n
\ee
we see that eq.~(\ref{hypoImnl}) holds. This proves the assertion.$\
\  \Box $

\begin{lemma}\label{hphimul}
For all $a,b \in \bC $, we have
\be \label{eamul}
:\exp \{ a\Phi \} :_\g \cdot :\exp \{ b\Phi \} :_\g =
\exp \{ \g ab \} \, :\exp \{ (a+b) \Phi \} :_\g
\ee
and, for $4\g^2ab<1$, $\cL_q := (1+2\g q)^{-1}$,
\be \label{ea2mul}
:\exp \{ a\Phi^2 \} :_\g \cdot :\exp \{ b\Phi^2 \} :_\g =
(\cL_a \cL_b \cL_{-(a\cL_a +b\cL_b)})^{\half}
:\exp \{ (a\cL_a +b\cL_b) \cL_{-(a\cL_a +b\cL_b)} \Phi^2 \} :_\g
\ee
\end{lemma}
\begin{lemma}\label{eaphimul}
For $a,b \in \bC $ define
\be
f_m(a) := \exp \{ \frac{a^2}{2}\} \, \frac{a^m}{m!}, \qquad
g_m(a) := \exp \{ -\frac{a^2}{2}\} \, \frac{(-a^2)^m}{(2m)!}, \qquad
h_m(a) := \exp \{ \frac{a^2}{2}\} \, \frac{a^{2m}}{(2m)!}.
\ee
Then we have
\be \label{pmul1}
\sum_{m,n:\atop |m-n|\le l \le m+n} f_m(a)\, f_n(b) \, C_l^{mn} =
  f_l(a+b),
\ee
for all $a,b\in \bC $;
\be \label{pmul2}
\sum_{m,n:\atop |m-n|\le l \le m+n} g_m(a)\, g_n(b) \, \cC_l^{mn} =
  g_l(a+b),
\ee
for all $a,b\in \bR $; and
\be \label{pmul3}
\sum_{m,n:\atop |m-n|\le l \le m+n} h_m(a)\, h_n(b) \, \cC_l^{mn} =
  \frac{1}{2}\, [h_l(a+b) + h_l(a-b)],
\ee
for all $a,b\in \bC $.
Furthermore, we have for all $a,b \in \bR $,
\be \label{cmulrel}
\sum_{m,n:\atop |m-n|\le l \le m+n} \frac{a^mb^n}{m!n!} \cC_l^{mn} =
\cN (a,b) \frac{c(a,b)^l}{l!},
\ee
where
\be
\cN (a,b) := \frac{1}{\sqrt{1-4ab}},\qquad
c(a,b) := \frac{a+b-4ab}{1-4ab}.
\ee
Generally, we have for all $N\ge 2$, $a_1, \ldots ,a_n \in \bR $,
\be \label{gencmulrel}
\sum_{l_1,l_2,\ldots ,l_{N-2}\atop m_1,m_2,\ldots ,m_N}
\cC_{l_1}^{m_1m_2} \cC_{l_2}^{l_1m_3}
   \cdots \cC_l^{l_{N-2}m_N} \,
      \frac{a_1^{m_1} \cdots a_N^{m_N}}{m_1!\cdots m_N!} =
  \frac{1}{l!}\, \cN (a_1,\ldots ,a_N) \, c(a_1,\ldots ,a_N)^l,
\ee
where
\bea
\cN (a_1,\ldots ,a_N) &:=& (\cL_{a_1} \cdots \cL_{a_N} \cL_{-(a_1\cL_{a_1} +
   \ldots +a_N\cL_{a_N})} )^{\frac{1}{2}},
\nonu\\
c(a_1,\ldots ,a_N) &:=& (a_1\cL_{a_1} + \ldots +a_N\cL_{a_N})
  \cL_{-(a_1\cL_{a_1} + \ldots +a_N\cL_{a_N})}
\eea
and $\cL_q = (1+2q)^{-1}.$
Moreover,
\be \label{qwmul2}
\sum_{m,n:\atop |m-n|\le l \le m+n}
\frac{a^{m+n}}{m!n!} \cC_l^{mn} =
  \frac{1}{l!}\, \frac{1}{\sqrt{1-4a^2}}(\frac{2a}{1+2a})^l.
\ee
\end{lemma}
\underline{Proof:\,} Eq.~(\ref{eamul})
of Lemma \ref{hphimul} implies
\be \label{cgj}
\sum_{m,n:\atop |m-n|\le l \le m+n} \frac{a^mb^n}{m!n!} C_l^{mn} =
\frac{(a+b)^l}{l!} \, \exp \{ ab \}
\ee
{}From this equation follows eq.~(\ref{pmul1}).
Eq.~(\ref{pmul2}) can be implied by replacing $a,b\in \bR $ by $ia,ib$
in eq.~(\ref{cgj}) and taking the real part. Using eq.~(\ref{cgj})
and change the signs of $a$ and $b$
and add the four resulting equalities we obtain
\be \label{qwmul1}
\sum_{m,n:\atop |m-n|\le l \le m+n}
\frac{a^{2m}b^{2n}}{(2m)!(2n)!} \cC_l^{mn} =
\frac{1}{2(2l)!}\, [\exp \{ ab \} (a+b)^{2l} +
      \exp \{ -ab \} (a-b)^{2l} ].
\ee
This implies Eq.~(\ref{pmul3}). Eq.~(\ref{cmulrel}) follows from
eq.~(\ref{ea2mul}) of Lemma \ref{hphimul}. The generalization
eq.~(\ref{gencmulrel}) follows also immediately from
eq.~(\ref{ea2mul}) of Lemma \ref{hphimul}. Furthermore,
eq.~(\ref{qwmul1}) follows from eq.~(\ref{cmulrel}) if
we chance the signs of $a$ and $b$ and add the four resulting equalities.
Eq.~(\ref{qwmul2}) follows eq.~(\ref{cmulrel}) if we set $a=b$.
$\  \  \Box $

\abstand
For $C_l^{mn}$ there exists recursive equations.
\medskip\noindent
\begin{lemma} \label{lrec} We have for all $l,m,n \ge 0$
\be
lC_l^{mn} = mC_{l-1}^{m-1,n} + nC_{l-1}^{m,n-1}.
\ee
\end{lemma}
\vskip5mm\noindent
\underline{\sl Proof:\,} Set $\g =1.$ Partial integration gives
\be
\langle F(\Phi ) \, :\Phi^l:\rangle = \langle
   \frac{\partial }{\partial \Phi } F(\Phi )\, :\Phi^{l-1}: \rangle .
\ee
Then, we obtain,
\bea
\langle :\Phi^m: :\Phi^n: :\Phi^l:\rangle &=& m
\langle :\Phi^{m-1}: :\Phi^n: :\Phi^{l-1}:\rangle
\nonu\\ &+& n
\langle :\Phi^m: :\Phi^{n-1}: :\Phi^{l-1}:\rangle .
\eea
Using
\be
\langle :\Phi^m: :\Phi^n: :\Phi^l:\rangle = C_l^{mn}
\langle :\Phi^l: :\Phi^l:\rangle = C_l^{mn} \, l!,
\ee
we obtain
\be
C_l^{mn} \, l! = C_{l-1}^{m-1,n} m\, (l-1)! +
  C_{l-1}^{m,n-1} n\, (l-1)!.
\ee
This implies the assertion. $\  \  \Box $

\section{Norms and Scalar Products}
\label{NorSca}
When we follow a renormalization group flow, we only want to keep
track explicitly of relevant and marginal degrees of freedom.
We want to control remainders using norm estimates. It turns out
that the norms have to be well adjusted to the class of functions
obtained by RGTs. Roughly speaking we treat the large field problem
with the help of norms. From the Banach algebra point of view
we obtain a number of interesting estimates useful in a structural
analysis.

This section introduces some norms which will be useful in hierarchical
renormalization group studies. We define a norm $\Vert \cdot \Vert_\rho $
for the representations of the algebra by functions. Then we introduce
supremums norm $\Vert \cdot \Vert_\rho^{(\infty )}$ on $\bR^\infty $
which will be useful for $\epsilon $-expansions and other fixed point
methods. Furthermore, we define a norm $\Vert \cdot \Vert_{\rho ,u,\delta }$.
We show that the norm $\Vert \cdot \Vert_{\rho ,0,0}$ of
the UV fixed point $Z_{UV}$ is infinite.
It was shown by Koch and Wittwer \cite{KW91} that the
norm $\Vert \cdot \Vert_{\rho ,0,0}$ of the
infrared fixed point $Z_{IR}$ is finite for $\beta =L^{-1-\frac{d}{2}}$,
in $d=3$ dimensions.
Furthermore, we define a norm
$\Vert \cdot \Vert_{(\g )}$ from a scalar product $<\cdot ,\cdot >$.
This norm is useful for deriving an upper bound for the (double-well)
infrared fixed points.
\subsection{The norm $\Vert \cdot \Vert_\rho $}
Our first norm is a supremum norm with a large field regulator.
Let $F$ be a real analytic function and $\rho \in \bR $.
Define a norm by
\be
\Vert F\Vert_\rho := \sup_{\Phi \in  \bR } \exp
            \{ -\rho \Phi^2 \} \, | F(\Phi )|.
\ee
Denote the vector space of all real analytic functions $Z$ with finite
norm $\Vert Z\Vert_\rho < \infty $ by $\cB_\rho $.
Obviously,
\be
\cB_{\rho_1} \subseteq \cB_{\rho_2}, \qquad \hbox{if}\ \rho_1 \le \rho_2
\ee
and
\be
\Vert A\cdot B\Vert_{\rho_1+\rho_2} \le \Vert A\Vert_{\rho_1} \cdot
                                       \Vert B\Vert_{\rho_2}.
\ee
Lemma \ref{lemma1.1} implies
\be
\Vert \cR_{\g ,L^d}^\beta (F)
\Vert_{\frac{L^d \beta^2\rho}{1-2L^d\g \rho }} \le
(1-2\g \rho )^{-\frac{L^d}{2}}\Vert F\Vert_\rho^{L^d}
\ee
for $\rho < \frac{1}{2\g }.$
Therefore, we see that
\be
\cR_{\g ,L^d}^\beta (\cB_\rho ) \subseteq \cB_{F(\rho )},
\ee
where
\be
F(\rho ) := \frac{L^d \beta^2\rho}{1-2L^d\g \rho }.
\ee
For
\be
\rho^*:= \frac{1-L^d\beta^2}{2L^d\g },
\ee
we have $F(\rho^*)=\rho^*$. Since $\cB_{\rho_1} \subseteq \cB_{\rho_2}$
if $\rho_1 \le \rho_2$ and $F(\rho ) \le \rho $ if $\rho \le \rho^*$,
we obtain
\be
\cR_{\g ,L^d}^\beta (\cB_\rho ) \subseteq \cB_\rho ,
\ee
for $\rho \le \rho^*$.
Thus we see that the normed space $\cB_{\rho^*}$ is invariant under RGTs.
For the special case $\beta = L^{1-\frac{d}{2}}$,
we have $\rho^* =\rho^*_{HT}$, where
\be
\rho^*_{HT} := -\frac{L^2-1}{2\g L^d}= \frac{1-L^d\beta^2}{2\g L^d}.
\ee
For the special case $\beta^{'}=L^{-1-\frac{d}{2}}$,
we have $\rho^* =\rho^*_{IR}$,
\be
\rho^*_{IR} := \frac{1-L^{-2}}{2\g L^d}=\frac{1-L^d\beta^{'2}}{2\g L^d}.
\ee
\begin{lemma}\label{banast}
Suppose that $\rho \in \bR_+$, $\rho < \frac{1}{4\g }$, $2\beta^2 <1$.
The product $\ast $ defines a mapping
$\ast :\, \cB_\rho^2 \rightarrow \cB_{\rho^{'}}$, where
\be
\rho^{'} := \frac{2\rho \beta^2}{1-4\g \rho},
\ee
and $\cB_{\rho^{'}}\subseteq \cB_{\rho}$ if
$\rho \le \rho^*_{\ast } := \frac{1-2\beta^2}{4\g }$.
\end{lemma}
\underline{\sl Proof:\,} Suppose that $A,B\in \cB_\rho $. Then
\be
:A:_\g \cdot :B:_\g \in \cB_{2\rho }.
\ee
This implies
\be
E_\g (A\cdot B) \in \cB_{\frac{2\rho }{1-4\g \rho }}.
\ee
Furthermore,
\be \label{assd}
\cS_\beta E_\g (A\cdot B)\in \cB_{\frac{2\beta^2\rho }{1-4\g \rho }}.
\ee
$\rho \le \rho^*_{\ast } $
implies $\rho^{'} \le \rho $.
Thus by eq.~(\ref{assd}) $A\ast B \in \cB_{\rho^{'}}.\  \  \  \Box $

\abstand

In the Banach algebra picture the following norm takes best care
of the particular form of the multiplication table.
Define a norm $\Vert \cdot \Vert_\rho^{(\infty )}$ on $\bR^\infty $ by
\be
\Vert a\Vert_\rho^{(\infty )}:= \sup_{l:\, l\in \bN } (|a_l|\, \rho^l\,
  l!).
\ee
Define a function $G :\, \bR_+ \rightarrow \bR_+ $ by
\be
G(\rho ) := 2\beta^2\rho - 2.
\ee
For $a,b \in \cR^\infty $ define
\be
(a\times_\beta b)_l := \beta^{2l}\, \sum_{m,n:\atop |m-n| \le l \le m+n}
\cC_l^{mn}\, a_m \, b_n.
\ee
\begin{lemma} \label{lsupbou} For $\rho \in \bR_+$,
$a,b \in \bR^\infty $ we have
\be
\Vert a\times_\beta b\Vert_\rho^{(\infty )}
   \le \frac{1}{\sqrt{1-(\frac{2}{\rho})^2}}\,
    \Vert a\Vert_{G(\rho )}^{(\infty )}\,
       \Vert b\Vert_{G(\rho )}^{(\infty )} .
\ee
\end{lemma}
\vskip5mm\noindent
\underline{\sl Proof:\,} We have
\be
|(a\times_\beta b)_l| \rho^l\,
  l! \le (\beta^2 \rho )^l \,
     \sum_{m,n\in \bN \colon \atop |m-n| \le l \le m+n}
        \frac{(\frac{1}{\rho })^m\, (\frac{1}{\rho })^n}{m!\, n!}
          \, l!\,
          \cC_l^{mn} \, \Vert a\Vert_\rho^{(\infty )} \,
                 \Vert b\Vert_\rho^{(\infty )} .
\ee
Using Lemma \ref{eaphimul}, eq.~(\ref{qwmul2}) we obtain
\be
|(a\times_\beta b)_l| \rho^l\,
  l! \le (\beta^2 \rho )^l \,
  \frac{1}{\sqrt{1-(\frac{2}{\rho})^2}}\,
    (\frac{\frac{2}{\rho}}{1+\frac{2}{\rho}})^l
\, \Vert a\Vert_\rho^{(\infty )} \,
                 \Vert b\Vert_\rho^{(\infty )} .
\ee
This implies
\be
|(a\times_\beta b)_l| (G^{-1}(\rho ))^l\,
  l! \le
  \frac{1}{\sqrt{1-(\frac{2}{\rho})^2}}\,
\Vert a\Vert_\rho^{(\infty )} \,
                 \Vert b\Vert_\rho^{(\infty )}.
\ee
Replacing $\rho $ by $G(\rho )$ we obtain the assertion. $\  \  \Box $

\abstand
The function $G :\, \bR_+ \rightarrow \bR_+ $ has a fixed point
\be
\rho^*_\infty := \frac{2}{2 \beta^2-1}.
\ee
We have $G(\rho ) \le \rho $ for $\rho \le \rho^*_\infty $ and
$G(\rho ) \ge \rho $ for $\rho \ge \rho^*_\infty $.
This and Lemma \ref{lsupbou} proves the following lemma.

\begin{lemma} \label{lVVertaln}
For $\rho \le \rho^*_\infty $,
$(\bR_{\rho }^\infty ,\times_\beta )$
is a Banach algebra and for $\beta \in (\frac{1}{\sqrt{2}},1)$ we have
\be
\Vert a\times_\beta b\Vert_{\rho^*_\infty}^{(\infty )}
   \le \frac{1}{2\beta \sqrt{1-\beta^2}}\,
    \Vert a\Vert_{\rho^*_\infty}^{(\infty )}\,
         \Vert b\Vert_{\rho^*_\infty}^{(\infty )} .
\ee
\end{lemma}
Suppose that for $a=\sum_m a_m\, e_{2m},\  e_{2m} = :\Phi^{2m}:_1$,
the weight $\omega (a) $ of $a$ is defined by eq.~(\ref{defome}).

\begin{lemma} If $\Vert a\Vert_{\rho_\infty^*}^{(\infty )} < \infty $
then the series expansion of $\omega (a)$ is convergent for all
$\beta =2^{\frac{2-d}{2d}},$ $d>2$, and
\be
|\omega (a)| \le
\frac{\Vert a\Vert_{\rho^*_\infty}^{(\infty )}}{4(1-\beta^2)} .
\ee
\end{lemma}
\underline{Proof:\,} $\Vert a\Vert_{\rho_\infty^*}^{(\infty )} < \infty $
implies
\be
|a_l| < \frac{C}{l!}\, (\rho^*_\infty )^{-l}, \qquad C:=
\Vert a\Vert_{\rho^*_\infty}^{(\infty )}.
\ee
Thus, using $\frac{(2m)!}{m!^2} \le \frac{1}{2}\, 4^m$,
\bea
|\omega (a)| &\le &
\sum_m \frac{(2m)!}{m!} \, 2^{-m} |a_m| \le  \sum_m
\frac{(2m)!}{m!} \, \frac{C}{(2\rho_\infty^*)^m}
\nonu\\  &\le &
\frac{C}{2} \sum_m (2\beta^2-1)^m =
\frac{C}{4(1-\beta^2)} . \  \  \Box
\eea
\subsection{The norm $\Vert \cdot \Vert_{\rho ,\alpha ,\delta }$}
Define a norm $\Vert \cdot \Vert_{\rho ,\alpha ,\delta }$ on $\bR^\infty $ by
\be
\Vert (a_0,a_1,a_2,\dots )\Vert_{\rho ,\alpha ,\delta } := \sum_{n=0}^\infty
|a_n| n!^\alpha (n+1)^\delta \rho^n,
\ee
for all $\alpha \in \bR $, $\rho \in \bR_+$, $a:=(a_0,a_1,a_2, \ldots ) \in
\bR^\infty $. Then
\be
\bR_{\rho ,\alpha ,\delta }^\infty :=
(\{ a\in \bR^\infty |\, \Vert a\Vert_{\rho ,\alpha ,\delta }
< \infty \} ,\Vert \cdot \Vert_{\rho ,\alpha ,\delta })
\ee
is a Banach space. We will use the notation $\Vert \cdot
\Vert_{\rho ,\alpha ,\delta } = \Vert \cdot \Vert_{\rho }$ for $\alpha =
\delta =0$.
For $a,b \in \bR^\infty $ define a product
\be
(a\ast_\beta b)_l := \sum_{m,n\in \bN \colon \atop |m-n| \le l \le m+n}
\beta^{l+m+n}\, A_l^{mn}\, a_m b_n ,
\ee
where
\be
A_l^{mn} := \sqrt{\frac{(2l)!}{(2m)! (2n)!}} \cS_l^{mn}.
\ee
Consider the function $g:\, \bR^\infty \rightarrow \cB $ defined by
\be
g(a)(\Phi ) := a(\Phi ) := \sum_{n:\, n\in \bN } (\frac{\beta }{\gamma })^n
            \frac{a_n}{\sqrt{(2n)!}} \Phi^{2n} .
\ee
$g$ is a $\ast_\beta $-homomorphism of $(\bR^\infty ,\ast_\beta )$
into $(\cB ,\ast_\beta )$, i.~e.
\be
g(a\ast_\beta b) (\Phi )= a(\Phi ) \ast_\beta b(\Phi ).
\ee
Define a function $G_\alpha :\, \bR_+ \rightarrow \bR_+ $ by
\be \label{gadef}
G_\alpha (\rho ) := 2^{1+\alpha }(\beta + \beta^2\rho ).
\ee
\begin{lemma} \label{lastbou} For $\alpha ,\rho \in \bR_+$,
$\delta \ge 0$ and $a,b \in \bR^\infty $, we have
\be
\Vert a \ast_\beta b \Vert_{\rho ,\alpha ,\delta } \le
                \Vert a
        \Vert_{G_\alpha (\rho ), \alpha ,\delta }\,
\Vert b \Vert_{G_\alpha (\rho ),\alpha ,\delta }.
\ee
\end{lemma}
\vskip5mm\noindent
\underline{\sl Proof:\,} Using Lemma \ref{cbouast} and $(m+n)! \le
2^{m+n} m!n!$ we obtain
\bea
\Vert a \ast_\beta b \Vert_{\rho ,\alpha ,\delta } &=&
\sum_l | \sum_{m,n:\atop |m-n|\le l \le m+n} \beta^{l+m+n} a_mb_n A_l^{mn}|
\, l!^\alpha \, (l+1)^{\delta }\, \rho^l
\nonu\\ &\le &
\sum_l \sum_{m,n:\atop |m-n|\le l \le m+n} |a_m||b_n| \beta^{l+m+n}
\sqrt{2(m+n) \choose 2m} {m+n\choose l}
\nonu\\ & &
2^{\alpha (m+n)} m!^\alpha
n!^\alpha (m+1)^\delta (n+1)^\delta \rho^l
\nonu\\ &\le &
\sum_l \sum_{m,n:\atop |m-n|\le l \le m+n} |a_m|m!^\alpha \, (m+1)^\delta
\, |b_n| n!^\alpha \, (n+1)^\delta \, (2\beta )^{m+n-l}
\nonu\\ & &
2^{\alpha (m+n)} {m+n\choose l} (2\beta^2\rho )^l
\nonu\\ &\le &
\sum_{m,n} |a_m|m!^\alpha \, (m+1)^\delta \, |b_n| n!^\alpha \,
(n+1)^\delta \,
[2^{\alpha +1}(\beta +\beta^2 \rho )]^{m+n}
\nonu\\ &=&
\Vert a \Vert_{G_\alpha (\rho ), \alpha ,\delta } \, \Vert b
\Vert_{G_\alpha (\rho ), \alpha ,\delta } .
\eea
This proves the assertion. $\  \  \ \Box $

\abstand
The function $G_\alpha :\, \bR_+ \rightarrow \bR_+ $ has a fixed point
\be
\rho^*_\ast := \frac{2^{1+\alpha } \beta}{1-2^{1+\alpha }\beta^2}.
\ee
We have $G_\alpha (\rho ) \le \rho $ for $\rho \ge \rho^*_\ast $ and
$G_\alpha (\rho ) \ge \rho $ for $\rho \le \rho^*_\ast $.
This and Lemma \ref{lastbou} proves the following Lemma.

\begin{lemma} For $\rho \ge \rho^*_\ast $ and
$\beta < 2^{-\frac{1+\alpha}{2}}$,
$(\bR_{\rho ,\alpha ,\delta }^\infty ,\ast )$
is a Banach algebra.
\end{lemma}
\begin{lemma} \label{nfac2} For all $n\ge 1$, we have
\be
\frac{(2n)!}{n!^2} =q_n\, 2^{2n},
\ee
where
\be \label{defqn}
q_n := \prod_{k=1}^n \frac{2k-1}{2k}, \qquad \frac{1}{\sqrt{\pi n}} \le
q_n \le \frac{1}{2} .
\ee
Furthermore,
\be
\lim_{n\rightarrow \infty } q_n \sqrt{n} = \frac{1}{\sqrt{\pi}}
\ee
and
\be
\sqrt{2n-1} \, q_n \le \sqrt{\frac{2}{\pi}} \exp \{ -\frac{1}{4n} \} .
\ee
\end{lemma}
\vskip5mm\noindent
\underline{\sl Proof:\,} Wallis' formula gives
\be
\sqrt{\frac{\pi}{2}} =\lim_{n\rightarrow \infty}
     \frac{2\cdot 4\cdot 6 \cdots (2n-2)}{3\cdot 5\cdot 7 \cdots (2n-1)}
        \sqrt{2n} .
\ee
Thus
\be
\sqrt{\frac{\pi}{2}} =\lim_{n\rightarrow \infty}
     \frac{2^2\cdot 4^2\cdot 6^2 \cdots (2n-2)^2}{(2n-1)!} \sqrt{2n}
     = \lim_{n\rightarrow \infty}
     \frac{2^2\cdot 4^2\cdot 6^2 \cdots 2n^2}{(2n)!\, 2n} \sqrt{2n}.
\ee
This implies
\be
\sqrt{\pi} = \lim_{n\rightarrow \infty} \frac{n!^2 2^{2n}}{(2n)!\, \sqrt{n}}
           = \lim_{n\rightarrow \infty} \frac{1}{q_n\, \sqrt{n}}.
\ee
We have
\bea
((2n-1)q_n^2)^{-1} &=& \frac{2^2}{1\cdot 3} \, \frac{4^2}{3\cdot 5}
         \frac{6^2}{5\cdot 7} \cdots \frac{(2n-2)^2}{(2n-3)\cdot
     (2n-1)} \nonu\\ &=&
 \frac{2^2}{2^2-1}\cdot  \frac{4^2}{4^2-1}\cdots
       \frac{(2n-2)^2}{(2n-2)^2-1} \nonu\\ &=&
     \frac{1}{1-\frac{1}{2^2}}\cdot  \frac{1}{1-\frac{1}{4^2}}\cdots
       \frac{1}{1-\frac{1}{(2n-2)^2}}.
\eea
This implies
\be
\frac{\lim_{n\rightarrow \infty }
  (2n-1)q_n^2}{(2n-1)q_n^2} = \frac{1}{1-\frac{1}{(2n)^2}}\cdot
      \frac{1}{1-\frac{1}{(2n+2)^2}}\cdots .
\ee
Therefore
\be
\frac{2}{\pi} \cdot \frac{1}{(2n-1)q_n^2} =  \frac{1}{1-\frac{1}{(2n)^2}}
   \cdot  \frac{1}{1-\frac{1}{(2n+2)^2}}\cdot \ldots .
\ee
Since
\be
\frac{1}{(2n)^2} + \frac{1}{(2n+2)^2} + \ldots  \ge \frac{1}{2n}
\ee
we obtain
\bea
(2n-1)q_n^2 &=& \frac{2}{\pi} \cdot (1-\frac{1}{(2n)^2})\cdot
          (1-\frac{1}{(2n+2)^2})\cdot \ldots
  \nonu\\ &=&
 \frac{2}{\pi} \exp \{ -\frac{1}{(2n)^2} - \frac{1}{(2n+2)^2} - \ldots \}
 \le \frac{2}{\pi} \, \exp \{ -\frac{1}{2n} \} .
\eea
This proves the assertion. $\  \  \ \Box $

\abstand
Recall that the constants of multiplication $\cS_l^{mn}$ arise
in the basis given by simple powers of fields. In terms of
normal ordered products we found the constant of multiplication
$\cC_l^{mn}.$ Similar norm estimates to those above can also be
proved in the normal ordered case.

For $a=(a_0,a_1,a_2, \ldots ),\, b=(b_0,b_1,b_2, \ldots ) \in
\bR^\infty $ define a product
\be
(a\twtimes_\beta b)_l := \sum_{m,n\in \bN \colon \atop |m-n| \le l \le m+n}
\beta^{l+m+n}\, B(l,m,n)\, a_m b_n  ,
\ee
where
\be
B(l,m,n):= \sqrt{\frac{(2l)!}{(2m)!\, (2n)!}} \, \cC_l^{mn}.
\ee

Consider the function $f:\, \bR^\infty \rightarrow B$ defined by
\be
f(a) := a(\Phi ) := \sum_{n:\, n\in \bN } (\frac{\beta }{\gamma })^n
            \frac{a_n}{\sqrt{(2n)!}} :\Phi^{2n}:_\g .
\ee
$f$ obeys
\be
f(a\twtimes_\beta b) = a(\Phi ) \times_\beta b(\Phi ).
\ee

Define a function $F_\alpha :\, \bR_+ \rightarrow \bR_+ $ by
\be
F_\alpha (\rho ) := 2^\alpha (\beta + 2\beta^2\rho ).
\ee
\begin{lemma} \label{lprodbou} For $\alpha \in \bR $, $\rho \in \bR_+$,
$\delta \ge 0$ and $a,b \in \bR^\infty $, we have
\be
\Vert a \twtimes_\beta b \Vert_{\rho ,\alpha ,\delta } \le
              \Vert a
        \Vert_{F_\alpha (\rho ), \alpha ,\delta }\,
\Vert b \Vert_{F_\alpha (\rho ),\alpha ,\delta }.
\ee
\end{lemma}
\vskip5mm\noindent
\underline{\sl Proof:\,} Using Lemma \ref{cbou} and $(m+n)! \le
2^{m+n} m!n!$ we obtain
\bea
\Vert a \twtimes_\beta b \Vert_{\rho ,\alpha ,\delta } &=&
\sum_l \  | \sum_{m,n:\atop |m-n|\le l \le m+n} a_mb_n B(l,m,n)|
\, l!^\alpha \, (l+1)^{\delta }\, \rho^l
\nonu\\ &\le &
\sum_l \sum_{m,n:\atop |m-n|\le l \le m+n} |a_m||b_n| \beta^{l+m+n}
\sqrt{2l \choose l} {m+n\choose l}
\nonu\\ & &
2^{\alpha (m+n)} m!^\alpha
n!^\alpha (m+1)^\delta (n+1)^\delta \rho^l
\nonu\\ &\le &
\sum_l \sum_{m,n:\atop |m-n|\le l \le m+n} |a_m|m!^\alpha \, (m+1)^\delta
\, |b_n| n!^\alpha \, (n+1)^\delta \, \beta^{m+n-l}
\nonu\\ & &
2^{\alpha (m+n)} {m+n\choose l} (2\beta^2\rho )^l
\nonu\\ &\le &
\sum_{m,n} |a_m|m!^\alpha \, (m+1)^\delta \, |b_n| n!^\alpha \,
(n+1)^\delta \,
[2^{\alpha (m+n)}(\beta +2\beta^2 \rho )]^{m+n}
\nonu\\ &=&
\Vert a \Vert_{F_\alpha (\rho ), \alpha ,\delta } \, \Vert b
\Vert_{F_\alpha (\rho ), \alpha ,\delta } .
\eea
This proves the assertion.$\  \  \ \Box $

\abstand
The function $F_\alpha :\, \bR_+ \rightarrow \bR_+ $ has a fixed point
\be
\rho^* := \frac{2^\alpha \beta}{1-2^{1+\alpha }\beta^2}.
\ee
We have $F_\alpha (\rho ) \le \rho $ for $\rho \ge \rho^*$ and
$F_\alpha (\rho ) \ge \rho $ for $\rho \le \rho^*$.
This and Lemma \ref{lprodbou} proves the following Lemma.
\begin{lemma} For $\rho \ge \rho^*$, and
$\beta < 2^{-\frac{1+\alpha}{2}}$
$(\bR_{\rho ,\alpha ,\delta }^\infty ,
\twtimes )$ is a Banach algebra.
\end{lemma}

In the following we will only consider the case $\alpha = \delta =0$
and write $\bR_{\rho }^\infty = \bR_{\rho ,0,0}^\infty .$
Suppose that for $a=\sum_m a_m\, e_{2m},\  e_{2m} = :\Phi^{2m}:_1$,
the weight $\omega (a) $ of $a$ is defined by eq.~(\ref{defome}).

\begin{lemma} If $\Vert a\Vert_{\rho^*} < \infty $
then the series expansion of $\omega (a)$ is convergent for all
$\beta =2^{-\frac{2+d}{2d}},$ $d>2$, and
\be
|\omega (a)| \le
\frac{\Vert a\Vert_{\rho^*}}{4(1-\beta^2)} .
\ee
\end{lemma}
\underline{Proof:\,} $\Vert a\Vert_{\rho^*} < \infty $
implies
\be
|a_l| < \frac{C}{\sqrt{(2l!)}}\, (1-2\beta^2)^l, \qquad C:=
\Vert a\Vert_{\rho^*}.
\ee
Thus, using $\frac{(2m)!}{m!^2} \le \frac{1}{2}\, 4^m$,
\bea
|\omega (a)| &\le &
C\, \sum_m \frac{\sqrt{(2m)!}}{2^m\, m!} (1-2\beta^2)^m\, \le \frac{C}{2}
 \sum_m (1-2\beta^2)^m
\nonu\\  &\le &
\frac{C}{4(1-\beta^2)} . \  \  \Box
\eea
\subsection{The norm $\Vert \cdot \Vert_{(\g )}$}
In the following we will write $\Vert \cdot \Vert_\rho := \Vert \cdot
\Vert_{\rho ,0,0}$.
For two functions $F, G:\, \bR \rightarrow \bR $ define a scalar product
by
\be
<F,G>_\g := \iGaug :F(\Phi ):_\g \, :G(\Phi ):_\g
\ee
and a norm
\be
\Vert F\Vert_{(\g )} := \sqrt{<F,F>_\g }.
\ee
Define a function $e_b:\, \bR \rightarrow \bR $ by
\be
e_b(\Phi ) := \exp \{ b\Phi \} .
\ee
\begin{lemma}\label{lscal1} For all $\Psi \in \bR $, we have
\be \label{sca1}
<F,e_\Psi >_\g = F(\g \Psi )
\ee
and
\be \label{sca2}
\Vert e_\Psi \Vert_{(\g )}^2 = \exp \{ \g \Psi^2 \} .
\ee
\end{lemma}
\underline{\sl Proof:\,}  Using integration by parts we obtain
\bea
<(\cdot )^n, e_\Psi >_\g &=&
  \iGaug :\Phi^n:_\g \, \exp \{ \Phi \Psi \} :_\g
\nonu\\ &=&
\g^n \iGaug :\frac{\partial^n}{\partial \Phi^n} \exp \{ \Phi \Psi \} :_\g
\nonu\\ &=&
(\g \Psi )^n.
\eea
This implies eq.~(\ref{sca1}). Eq.~(\ref{sca2}) follows since
\be
\Vert e_\Psi \Vert^2_{(\g )} = <e_\Psi ,e_\Psi >_{(\g )} =
  e_\Psi (\g \Psi ) = \exp \{ \g \Psi^2 \} .
\ee
This proves the assertion. $\  \  \ \Box $
\begin{lemma}\label{lscal2} For $f:= (f_0,f_1,\ldots ) \in
  \bR_{\frac{c\beta}{\g }}^\infty $, $c,\beta ,\g \in \bR $ define
\be
F(\Phi ) := \sum_{n=0}^\infty (\frac{\beta}{\g })^n
             \frac{f^n}{\sqrt{(2n)!}} \Phi^{2n}.
\ee
Then we have
\be
\Vert F\Vert_{(c)} \le \Vert f\Vert_{\frac{c\beta }{\g }}.
\ee
\end{lemma}
\underline{\sl Proof:\,} Consider the orthonormal basis
$\frac{c^n\Phi^{2n}}{\sqrt{(2n)!}}$ . Then we have
\bea
\Vert F\Vert_{(c)}^2 &=& \sum_{n=0}^\infty <F,\Phi^{2n}>_c\,
   <\Phi^{2n},F>_c\, \frac{c^{2n}}{(2n)!}
\nonu\\ &=&
     \sum_{n=0}^\infty f_n^2 (\frac{c\beta}{\g })^{2n}
  \le \Vert f\Vert_{\frac{c\beta}{\g }}^2.\  \  \Box
\eea
\begin{lemma}\label{lscal3} For $F:\bR \rightarrow \bR ,$
$c,\Psi \in \bR $, we have
\be
|F(\Psi )| \le \Vert F\Vert_{(c)} \, \exp \{ \frac{\Psi^2}{c} \} .
\ee
\end{lemma}
\underline{\sl Proof:\,} Lemma \ref{lscal1} yields
\be
F(\Psi ) = <F,e_{\frac{\Psi}{c}}>_c.
\ee
Thus
\be
|F(\Psi )| \le \Vert F\Vert_{(c)} \cdot \Vert e_{\frac{\Psi}{c}}\Vert_{(c)}.
\ee
Lemma \ref{lscal1}, eq.~(\ref{sca2}), yields the assertion.$\  \  \ \Box $

\abstand
The following Lemma presents an upper bound for the infrared
fixed points. This bound shows the large field
behaviour of the fixed points with finite norm. The infrared fixed point
in three dimensions belongs to the class of functions with finite norm
(cf.~\cite{KW91}).
\begin{lemma}
Suppose that $F(\Phi ) = \sum_{n=0}^\infty f_n \Phi ^{2n}$
and $\Vert f\Vert_\rho <\infty $ for some $\rho >\rho_*^* $.
If $f\ast_\beta f =f$ then there
exists a positive constant $K$ such that
\be
|F(\Psi )| \le \exp \{ K\Psi^{\frac{2d}{d+2}} \}
\ee
for all $\Psi \in \bR $.
\end{lemma}
\underline{Proof:\,} We have, using Lemma \ref{lastbou}, $\rho_*^* =
\frac{2\beta}{1-2\beta^2}$
\be \label{huhu5}
\Vert f\Vert_*^* \le \Vert f\Vert^2_{\rho_*^* +2\beta^2 \rho}.
\ee
Recall that we have set $\alpha =\delta =0$ and $\rho_*^*$ is the
fixed point of eq.~(\ref{gadef}) of the function $G_\alpha .$
Define a function $H_f:\, \bR_+ \rightarrow \bR_+ $ by
\be
H_f(\rho ) = \Vert f\Vert_{\rho_*^* +\rho }.
\ee
By eq.~(\ref{huhu5}) follows
\be \label{huhu6}
H_f(\rho ) \le H_f ((2\beta^2)^n\rho )^{2^n}
\ee
for all $n\in \bN $. Define for $q\in \bR_+$
\be
c_q := \sup_{\rho \in [1,\frac{1}{2\beta^2}]} \rho^{-q} \, \ln H_f(\rho ).
\ee
Then, we have
\be \label{huhu7}
H_f(\rho ) \le \exp \{ c_q\, \rho^q \}
\ee
for all $\rho \in [1,\frac{1}{2\beta^2}].$ Consider $\rho \in [1,\infty ).$
There exists $n\in \bN $ such that
\be
(2\beta^2)^n \rho \in [1,\frac{1}{2\beta^2}].
\ee
By eq.~(\ref{huhu6}) and eq.~(\ref{huhu7}) follows
\be \label{huhu8}
H_f (\rho ) \le \exp \{ c_q\, 2^n (2\beta^2)^{nq} \rho^q \} .
\ee
For $q=\frac{d}{2}$ we have
\be
2(2\beta^2)^q = 2\cdot 2^{-\frac{2q}{d}} =1.
\ee
Thus eq.~(\ref{huhu8}) implies
\be \label{huhu9}
\Vert f\Vert_{\rho_*^*+\rho} \le \exp \{ c_q\, \rho^q \}
\ee
for all $\rho \in [1,\infty ).$ Using Lemma \ref{lscal3}, we obtain
\be
|F(\Psi )| \le \exp \{ K\Psi^{\frac{2d}{d+2}} \}
\ee
and Lemma \ref{lscal2} yields
\be
|F(\Psi )| \le \Vert f\Vert_{\frac{c\beta}{\g }} \, \exp \{
   \frac{\Psi^2}{c} \} .
\ee
By eq.~(\ref{huhu9}) follows
\be
|F(\Psi )| \le \exp \{ c_q\, (\frac{c\beta}{\g }-\rho_*^*)^q +
   \frac{\Psi^2}{c} \}
\ee
for $\frac{c\beta}{\g } \ge 1.$ Define
\be \label{huhu10}
g(c) := c_q\, (\frac{c\beta}{\g }-\rho_*^*)^q +
   \frac{\Psi^2}{c} .
\ee
Let $x$ be defined such that
\be
c= \frac{\g}{\beta} \, (x+\rho_*^*).
\ee
Since $\rho_*^* \ge 1$, we have $\frac{c\beta}{\g }\ge 1$.
Thus eq.~(\ref{huhu10}) implies
\be \label{huhu11}
g(c) \le c_q\, x^q + \frac{\beta \Psi^2}{\g x} .
\ee
The minimal value of the rhs of inequality (\ref{huhu11}) reads
\be
x = (\frac{\beta \Psi^2}{\g qc_q})^{\frac{1}{q+1}}.
\ee
For this value of $x$ we get
\be
g(c) \le K(\beta ,\g ,d) \,  \Psi^{\frac{2d}{d+2}},
\ee
where
\be
K(\beta ,\g ,d) := c_q (\frac{\beta}{\g qc_q})^{\frac{d}{d+2}} +
\frac{\beta }{\g} (\frac{\g qc_q}{\beta})^{\frac{2}{d+2}}.
\ee
This proves the assertion. $\  \  \ \Box $

\abstand
The following Lemma gives a bound on functions with a quadratic
potential. The statement is that the norm is finite provided that the modulus
of the (negative) mass squared is sufficiently small. The
ultraviolet fixed point turns out to be on the border of the convergent
region.
\begin{lemma} \label{htbanast}
Consider $Q_c(\Phi ) := \exp \{ c\Phi^2 \} $ for $c\in \bR $.
Then $Q_c(\Phi )$
is contained in the Banach algebra $(\bR_{\rho^*_\ast }^\infty ,\ast )$,
iff $c<c_* := \frac{1-2\beta^2}{4\g }.$
Then we have, for $\beta = 2^{-\frac{2+d}{2d}}$ and $\rho^*_\ast =
\frac{2\beta }{1-2\beta^2},$
\be
\Vert Q_c\Vert_{\rho^*_\ast } \le 1 + (\frac{2}{\pi})^{\frac{1}{4}}
\frac{\frac{2c\g \rho^*_\ast }{\beta }}{1-\frac{2c\g \rho^*_\ast }{\beta }}.
\ee
Let $Z_{UV}$ be the UV fixed point of
the RG transformation $\cR_{\g ,2}^\beta $.
Then $Z_{UV}$
is not contained
in the Banach algebra $(\bR_{\rho^*_\ast }^\infty ,\ast )$.
\end{lemma}
\vskip5mm\noindent
\underline{\sl Proof:\,} Using the definition of
the UV fixed point
$Z_{UV}$ we obtain
\be
Z_{UV}(\Phi ) =
2^{-\frac{1}{d}} \exp \{ c_* \Phi^2 \} .
\ee
Consider
\be
Q_c=(a_0,a_1,a_2,\ldots ) \in  \bR_{\rho^*_\ast }^\infty  ,
\ee
where
\be
a_n :=(\frac{\g }{\beta })^n \sqrt{\frac{(2n)!}{n!^2}}
         c^n=(\frac{2\g c}{\beta })^n \, \sqrt{q_n}.
\ee
Then, using Lemma \ref{nfac2},
\be
a_n = (\frac{2\g c}{\beta })^n
\left ( (\pi n)^{-\frac{1}{4}} + r_n \right ),
\ee
where $r_n \ge 0$ and $\lim_{n\rightarrow \infty } r_n \sqrt{n} =0$.
Thus
\be
\Vert a\Vert_{\rho^*_\ast } = \sum_{n:\, n\in \bN }
  (\frac{2c\g \rho^*_\ast }{\beta })^n
\left ( (\pi n)^{-\frac{1}{4}} + r_n \right ).
\ee
The series is convergent if
\be
c < \frac{\beta}{2\g \rho^*_\ast } = \frac{1-2\beta^2}{4\g } = c_*
\ee
and divergent if
\be
c \ge \frac{\beta}{2\g \rho^*_\ast } = \frac{1-2\beta^2}{4\g } = c_*.
\ee
For the convergent case we have, using \ref{nfac2}
\be
\Vert a\Vert_{\rho^*_\ast } \le 1+\sum_{n:\, n\in \bN }
  (\frac{2c\g \rho^*_\ast }{\beta })^n
 (\frac{2}{\pi})^{\frac{1}{4}} \exp \{ -\frac{1}{8n} \}
(2n-1)^{-\frac{1}{4}}.
\ee
This implies
\be
\Vert a\Vert_{\rho^*_\ast } \le 1 + (\frac{2}{\pi})^{\frac{1}{4}}
     \frac{\frac{2c\g \rho^*_\ast }{\beta }}{1-\frac{2c\g \rho^*_\ast }{\beta
}}.
\ee
This proves the assertion.$\  \  \ \Box $
\begin{lemma} \label{htban}
Consider $Q_c(\Phi ) := \exp \{ c\Phi^2 \} $ for $c\in \bR $.
Then $Q_c(\Phi )$
is contained in the Banach algebra $(\bR_{\rho^* }^\infty ,\twtimes )$,
iff $c<\frac{\beta}{2\g \rho}.$ Then we have, for $\beta =
2^{-\frac{2+d}{2d}}$ and $\rho^* =
\frac{\beta }{1-2\beta^2},$
\be
\Vert Q_c\Vert_{\rho^*} \le 1 + (\frac{2}{\pi})^{\frac{1}{4}}
     \frac{\frac{2c\g \rho^* }{\beta }}{1-\frac{2c\g \rho^* }{\beta }}.
\ee
Let $Z_{UV}$
be the UV fixed point of
the RG transformation $\cR_{\g (1-\beta^2), 2}^\beta $.
Then $Z_{UV}$
is not contained
in the Banach algebra $(\bR_{\rho^* }^\infty ,\twtimes )$.
\end{lemma}
\vskip5mm\noindent
\underline{\sl Proof:\,} The UV fixed point
$Z_{UV}$ of the RG transformation $\cR_{\g (1-\beta^2), 2}^\beta $ reads
\be
Z_{UV}(\Phi ) =
2^{-\frac{1}{d}} \exp \{ c_* \Phi^2 \} ,
\ee
where
\be \label{cstdef}
c_* := \frac{1-2\beta^2}{4\g (1-\beta^2)}.
\ee
Consider
\be
Q_c=(a_0,a_1,a_2,\ldots ) \in  \bR_{\rho^* }^\infty  ,
\ee
where
\be
a_n = (\frac{\g }{\beta })^n \frac{\sqrt{(2n)!}}{n!} c^n.
\ee
Lemma \ref{nfac2} shows that
\be
a_n =(\frac{2\g c}{\beta })^n
\left ( (\pi n)^{-\frac{1}{4}} + r_n \right ),
\ee
where $r_n \ge 0$ and $\lim_{n\rightarrow \infty } r_n \sqrt{n} =0$.
Thus
\be
\Vert a\Vert_\rho =\sum_{n:\, n\in \bN }
  (\frac{2c\g \rho }{\beta })^n
\left ( (\pi n)^{-\frac{1}{4}} + r_n \right ).
\ee
The series is convergent, for $\rho =\rho^*$, if
\be
c < \frac{1-2\beta^2}{2\g }
\ee
and divergent if
\be
c \ge \frac{1-2\beta^2}{2\g }.
\ee
For the convergent case we have, using Lemma \ref{nfac2}
\be
\Vert a\Vert_{\rho^*}=1+\sum_{n:\, n\in \bN }
  (\frac{2c\g \rho^* }{\beta })^n
 (\frac{2}{\pi})^{\frac{1}{4}} \exp \{ -\frac{1}{8n} \}
(2n-1)^{-\frac{1}{4}}.
\ee
This implies
\be
\Vert a\Vert_{\rho^*} \le 1 + (\frac{2}{\pi})^{\frac{1}{4}}
     \frac{\frac{2c\g \rho^* }{\beta }}{1-\frac{2c\g \rho^* }{\beta }}.
\ee
Since
\be
:\exp \{ c_*\Phi^2 \} :_\g = \cL_{c_*}^{\half} \exp \{ \frac{c_*}{1+2\g c_*}
  \Phi^2 \} ,
\ee
we see that $f^{-1}(Z_{UV})
\in \bR_{\rho^*}^\infty $ iff
\be
\frac{c_*}{1-2\g c_*} < \frac{1-2\beta^2}{2\g }.
\ee
This is equivalent to
\be
c_* < \frac{1-2\beta^2}{4\g (1-\beta^2)}.
\ee
By eq.~(\ref{cstdef}) this is not valid. This proves the assertion.
$\  \  \ \Box $

\abstand
Let us consider an approximate RGT where we only keep track of the first $N$
terms in the expansion
\be
A(\Phi ) = \sum_{n=0}^\infty \frac{a_n}{\sqrt{(2n)!}} \,
\frac{\Phi^{2n}}{\g^n}.
\ee
This leads us to define the following truncated product
depending on the order $N$ of truncation.

For $a,b \in \bR^\infty $ and $N\in \bN $ define a product
\be
(a\bullet_\beta b)_l := \sum_{m,n\in \bN \colon \atop l \le m+n}
a_m b_n  A_N(l,m,n),
\ee
where $A_N(l,m,n):=A(l,m,n) $ if $l\ge N$ and $=0$ if $l < N$.

\begin{lemma} \label{lbullbou} For $\alpha ,\rho \in \bR_+$,
$\delta \ge 0$, $N\in \bN $, $q_N$ defined by eq.~(\ref{defqn})
and $a,b \in \bR^\infty $, we have
\be
\Vert a \bullet_\beta b \Vert_{\rho ,\alpha ,\delta } \le
                \sqrt{q_N} \, \Vert a
        \Vert_{G_\alpha (\rho ), \alpha ,\delta }\,
\Vert b \Vert_{G_\alpha (\rho ),\alpha ,\delta }.
\ee
\end{lemma}

Suppose that $\omega :\, \cB \rightarrow \bC $ is a weight of
the algebra $(\cB ,\times_1).$ For $z\in \cB $ define $p(z) \in \bR_+$ by
\be
p(z) := |\omega (z)|.
\ee
$p$ is a pre-norm of $\cB $ which satisfies
\be
p(x\times_1 y) =p(x)\, p(y).
\ee
{}From a pre-norm we obtain a norm by the following canonical
construction.
Define an equivalence relation on $\cB $ by $x\sim y$ iff $x-y \in
\ker \omega .$ Then we conclude that $p$ is a norm on $\cB /\sim .$

Define
\bea
u_m &:=& \frac{e_{2m}}{\sqrt{(2m)!}} = \frac{:\Phi^{2m}:_{\g }}{\g^m
         \sqrt{(2m)!}},
\nonu\\
v_m &:=& \frac{f_{2m}}{\sqrt{(2m)!}} = \frac{\Phi^{2m}}{\g^m
         \sqrt{(2m)!}}.
\eea
We have
\be
v_m = \sum_{n=0}^m U_{mn} u_n,
\ee
where
\be
U_{mn} = U_{mn}(1), \qquad U_{mn}(\alpha ) := \sqrt{\frac{(2n)!}{(2m)!}}
          E_{2m,2n}(\alpha ) = \sqrt{\frac{(2m)!}{(2n)!}}
              \frac{\alpha^{m-n}}{(m-n)!}.
\ee
\begin{lemma} \label{Uesti} For all $m,n\in \bN ,$ $m\ge n$ we have
\be
|U_{mn}(\alpha )| \le {m \choose n} (2\alpha )^{m-n}.
\ee
\end{lemma}
\vskip5mm\noindent
\underline{\sl Proof:\,} We have, for $m\ge n$,
\be
|U_{mn}(\alpha )| \le {m \choose n} \sqrt{\frac{(2m)!}{m!^2}}
              (\sqrt{\frac{(2n)!}{n!^2}})^{-1} \,  \alpha^{m-n} =
   {m \choose n} \sqrt{\frac{q_m}{q_n}} (2\alpha )^{m-n} \le
      {m \choose n} (2\alpha )^{m-n}.
\ee
This proves the assertion. $\  \  \ \Box $

\begin{lemma} Consider for $\delta \in \bR_+$
\be
G(\Phi ) =\sum_m a_m u_m,\qquad  :G:_\delta (\Phi ) =\sum_n b_n u_n.
\ee
We have
\be
b_n = \sum_{m:\, m\ge n} a_m U_{mn} (-\frac{\delta }{\g })
\ee
and
\be
\Vert b \Vert_\rho \le \Vert a \Vert_{\rho +2\frac{\delta }{\g }}.
\ee
\end{lemma}
\vskip5mm\noindent
\underline{\sl Proof:\,} We have
\bea
\Vert b \Vert_\rho &=& \sum_n |b_n| \rho^n = \sum_n| \sum_{m:\, m\ge n}
  a_m U_{mn}(-\frac{\delta }{\g })| \rho^n \le
  \nonu\\ &\le &
  \sum_m| \sum_{n=0}^m U_{mn}(-\frac{\delta }{\g })| |a_m| \rho^n \le
   \sum_m \sum_{n=0}^m  {m\choose n} (2\frac{\delta }{\g })^{m-n}
    \rho^n |a_m| \le
    \nonu\\ &\le &
    \sum_m |a_m| (\rho +2\frac{\delta }{\g })^m =
        \Vert a\Vert_{\rho +2\frac{\delta }{\g }} .
\eea
This proves the assertion. $\  \  \ \Box $

\section{Twodimensional Fixed Points}
\label{AssPer}
In this section the case $\beta =1$ will be studied.
Because of the singularity of the RGT $\cR_{\frac{\g }{1-\beta^2},2}^\beta $
at $\beta =1$ the $\times_1$-multiplication cannot be used for the study
of RG fixed points. Therefore we will use the multiplication $\ast $
instead. We will show that there exists a continuum of
periodic fixed points in this case.

The $\ast $-multiplication of two exponential functions is
\be
\exp \{ a\Phi \} \ast \exp \{ b\Phi \} = \exp \{ \frac{\g }{2}(a+b)^2\} \,
   \exp \{ (a+b)\Phi \} .
\ee
Define, for $m\in \bZ $, $q\in \bR $,
\be
u_m := \exp \{ 2\pi im \frac{q}{\sqrt{\g }}\Phi \} .
\ee
Then
\be \label{twomul}
u_m \ast u_n = \exp \{ -2\pi^2q^2 (m+n)^2\} \, u_{m+n}.
\ee
We see that the multiplication rule
eq.~(\ref{twomul}) defines a nonassociative algebra, since
\be
(u_m \ast u_n)\ast u_l = \exp \{ -2\pi^2q^2 [(m+n)^2 +(m+n+l)^2]\} \,
u_{l+m+n} \ne u_m \ast (u_n \ast u_l).
\ee
But the algebra defined by eq.~(\ref{twomul}) does define a power algebra.

\noindent
For
\be
A:= \sum_{m:\, m\in \bZ } a_m u_m ,\qquad B:=
                  \sum_{n:\, n\in \bZ } b_n u_n,
\ee
the $\ast $-multiplication of $A$ and $B$ yields
\be
A \ast B = \sum_{l:\, l\in \bZ } c_l u_l,
\ee
where
\be
c_l := \exp \{ -2\pi^2q^2 l^2\} \, \sum_{m,n\in \bZ : \atop m+n=l} a_m b_n.
\ee
Thus, the fixed point equation
\be \label{fix2eq}
A\ast A = A
\ee
is equivalent to
\be \label{perfix}
a_l = \exp \{ -2\pi^2q^2 l^2\} \, \sum_{m,n\in \bZ : \atop m+n=l} a_m a_n.
\ee
The solution of the fixed point eq.~(\ref{fix2eq}) can be expressed
in terms of the Theta-functions.

Define Theta-functions $\Theta_2(z,q)$ and $\Theta_3(z,q)$,
for $|q|<1,$ $z\in \bC $, by
\bea
\Theta_2(z,q) &:=& \sum_{n:\, n\in \bZ } q^{n^2} \, \exp \{ 2niz\}
=1+2\sum_{n:\, n\in \bN } q^{n^2} \, \cos \{ 2nz\} ,
\nonu\\
\Theta_3(z,q) &:=& -\sum_{n:\, n\in \bZ +\frac{1}{2}} q^{n^2} \,
\exp \{ 2niz\} = -2\sum_{n:\, n\in \bN } q^{(n+\frac{1}{2})^2}
\, \cos \{ (2n+1)z\}.
\eea
Let us remark that $\Theta_2$ and $\Theta_3$ are even functions in $z$.
Furthermore, define
\bea
\Theta_+(q) &:=& \Theta_3(0,q^4) =\sum_{n:\, n\in 2\bZ } q^{n^2} ,
\nonu\\
\Theta_-(q) &:=& -\Theta_2(0,q^4)= -\sum_{n:\, n\in 2\bZ +1} q^{n^2} ,
\nonu\\
\Theta_{1/2}(q) &:=& \sum_{n:\, n\in 2\bZ +\frac{1}{2}} q^{n^2} .
\eea
\begin{lemma} \label{exsol}
A solution of eq.~(\ref{perfix}) is given by
\be
a_l := \left\{ \begin{array}{r@{\quad:\quad}l}
           \cN_1\,  p^{l^2} & l\ \mbox{even} \\
           \cN_2\,  p^{l^2} & l\ \mbox{odd,}
               \end{array} \right.
\ee
where
\be
p:=\exp \{ -4\pi^2q^2\} ,\qquad \cN_1:=\frac{1}{\Theta_{1/2}(p^2)},
\qquad \cN_2:=\frac{\Theta_{1/2}(p^2)-\Theta_{+}(p^2)}{\Theta_{-}(p^2)
\, \Theta_{1/2}^2(p^2)}.
\ee
\end{lemma}
\underline{\sl Proof:\,} For $l$ even we have
\be
\sum_{m:\, m\in \bZ } a_{l-m} a_m = \cN_1^2 \sum_{m:\, m\in 2\bZ }
  p^{(l-m)^2+m^2} + \cN_2^2 \sum_{m:\, m\in 2\bZ +1}
  p^{(l-m)^2+m^2}
\ee
and for $l$ odd
\be
\sum_{m:\, m\in \bZ } a_{l-m} a_m = \cN_1 \cN_2\, [\sum_{m:\, m\in 2\bZ }
  p^{(l-m)^2+m^2} + \sum_{m:\, m\in 2\bZ +1}
  p^{(l-m)^2+m^2}].
\ee
Shift $m\rightarrow m+\frac{l}{2}$ we obtain, for $l$ even
\be
\sum_{m:\, m\in 2\bZ } p^{(l-m)^2+m^2} =
\left\{ \begin{array}{r@{\quad:\quad}l}
           \Theta_+(p^2)\,  p^{\frac{l^2}{2}} & \frac{l}{2}\ \mbox{even} \\
           \Theta_-(p^2) \, p^{\frac{l^2}{2}} & \frac{l}{2}\ \mbox{odd,}
               \end{array} \right.
\ee
and
\be
\sum_{m:\, m\in 2\bZ +1} p^{(l-m)^2+m^2} =
\left\{ \begin{array}{r@{\quad:\quad}l}
           \Theta_-(p^2)\, p^{\frac{l^2}{2}} & \frac{l}{2}\ \mbox{even} \\
           \Theta_+(p^2)\, p^{\frac{l^2}{2}} & \frac{l}{2}\ \mbox{odd.}
               \end{array} \right.
\ee
For $l$ odd we obtain
\be
\sum_{m:\, m\in 2\bZ } p^{(l-m)^2+m^2} = p^{\frac{l^2}{2}}
  \sum_{m:\, m\in 2\bZ +1} p^{2m^2} = \Theta_{1/2}(p^2) \, p^{\frac{l^2}{2}}
\ee
and
\be
\sum_{m:\, m\in 2\bZ +1} p^{(l-m)^2+m^2} = p^{\frac{l^2}{2}}
  \sum_{m:\, m\in 2\bZ } p^{2m^2} = \Theta_{1/2}(p^2)\, p^{\frac{l^2}{2}}.
\ee
Therefore, we get
\be \label{sumst}
\sum_{m:\, m\in \bZ } a_{l-m} a_m =
   \left\{ \begin{array}{r@{\quad:\quad}l}
           p^{l^2}\, [\cN_1^2 \Theta_+(p^2) + \cN_2^2 \Theta_-(p^2)] &
       l\ \mbox{even} \\
           p^{l^2}\, \cN_1 \cN_2 \, \Theta_{1/2}(p^2) & l\ \mbox{odd,}
               \end{array} \right.
\ee
Insertion of eq.~(\ref{sumst}) into eq.~(\ref{perfix}) yields
\bea \label{norequ}
\cN_1 &=& \cN_1^2 \Theta_+ + \cN_2^2 \Theta_-
\nonu\\
\cN_2 &=& \cN_1 \cN_2 \, \Theta_{1/2}.
\eea
Eq.~(\ref{norequ}) is fulfilled by definition of $\cN_1$ and $\cN_2$.
This proves the assertion. \hfill $\Box $

\abstand
Define
\be
Z_*(\Phi ) = \sum_{l:\, l\in \bZ } a_l \, \exp \{ 2\pi i l
  \frac{q}{\sqrt{\g}} \Phi \} ,
\ee
where $a_l$ is defined in Lemma \ref{exsol}. Then we have
\be
Z_*(\Phi ) = \cN_1 \sum_{l:\, l\in 2\bZ } p^{l^2} \, \exp \{ 2\pi i l
  \frac{q}{\sqrt{\g}} \Phi \}
+\cN_2 \sum_{l:\, l\in 2\bZ +1} p^{l^2} \, \exp \{ 2\pi i l
  \frac{q}{\sqrt{\g}} \Phi \}
\ee
Using the definitions of the Theta-functions we get,
for $p:= \exp \{ -4\pi^2 q^2 \} $, the following solutions of the
fixed point equation
\be \label{fixsol}
Z_*(\Phi ) =\frac{\Theta_3(\frac{4\pi q}{\sqrt{\g }}\Phi ,
p)}{\Theta_{1/2}(p^2)} -
  \frac{\Theta_2(\frac{4\pi q}{\sqrt{\g }}\Phi , p)
(\Theta_{1/2}(p^2)-\Theta_{+}(p^2))}{\Theta_{-}(p^2)\, \Theta_{1/2}^2(p^2)}.
\ee
\begin{lemma} $Z_*^{(q)}(\Phi )$ defined by eq.~(\ref{fixsol}) is an even
periodic solution
of $\cR_{\g ,2}^{\beta =1} (Z) =Z$, for all $q\in \bR ,$ $|q|<1$.
\end{lemma}
In the following we will consider the algebra of even periodic functions.
Suppose that
\be
u_{-m} = u_m .
\ee
Then
\be
u_m \ast u_n =  g(m+n) u_{m+n},
\ee
where
\be
g(k) := \exp \{ -2\pi^2 q^2 k^2\} .
\ee
Define
\be
o_m := \cosh \{ 2\pi \frac{q}{\g } m \Phi \} .
\ee
Then, we have
\be
o_m \ast o_n = \frac{1}{2} \left ( g(m+n) o_{m+n} +
                                  g(m-n) o_{m-n}\right ) .
\ee
For
\be
A= \sum_{m\in \bN } a_m o_m
\ee
the fixed point equation $A\ast A = A$ is equivalent to
\be
a_l = \frac{1}{2} \left (\sum_{m,n\in \bN : \atop m+n=l} g(m+n)
     a_m a_n + \sum_{m,n\in \bN : \atop m-n=l} g(m-n)
               a_m a_n \right ) .
\ee

\section{Gauss-Hermite-Formula for Constants of Multiplication}
\label{GauHer}
In this section we want to find a coordinate transformation
such that the constants of multiplication $B_l^{mn}$ become
diagonal. Diagonality means here that $B_l^{mn}=0$ if $m\ne n.$

In the following we will explain the Gaussian method for interpolating
integrals. For two continuous functions $f,g$ define the scalar product
\be
(f,g):= \int_{-\infty}^\infty dx\, \exp \{ -x^2 \} f(x) g(x) .
\ee
Hermite polynomials ${\rm H}_n(x)$ are orthogonal with respect to this
scalar product
\be
({\rm H}_m,{\rm H}_n) = ({\rm H}_m,{\rm H}_m) \, \delta_{m,n}.
\ee
The generating function $S(\Phi ,z)$ for Hermite polynomials reads
\be
S(\Phi ,z) = \sum_{n=0}^\infty \frac{{\rm H}_n(\Phi )}{n!} z^n =
\exp \{ -z^2 +2z\Phi \} .
\ee
Therefore
\be
:\exp \{ z\Phi \} :_\g = \exp \{ -\frac{\g}{2} z^2 +z\Phi \} =
  S(\frac{\Phi}{\sqrt{2\g}},\sqrt{\frac{\g}{2}} z).
\ee
Thus
\be \label{wickherm}
:\Phi^n:_\g = (\frac{\g }{2})^{\frac{n}{2}} \,
                 {\rm H}_n(\frac{\Phi}{\sqrt{2\g}}).
\ee
\begin{theorem} \label{GauHerm}
Let $x_1,x_2,\ldots ,x_n$ be the zeroes of ${\rm H}_n(x)$ and
$w_1,\ldots ,w_n$ be the solutions of
\be
\sum_{i=1}^n {\rm H}_k(x_i) w_i = \left \{ \begin{array}{r@{\quad:\quad}l}
               ({\rm H}_0,{\rm H}_0), & k=0 \\ 0, & k\in \{ 1,\ldots ,n-1\} .
                 \end{array} \right.
\ee
If $P$ is a polynom with grade $\le 2n-1$, we have
\be
\int_{-\infty }^\infty dx\, e^{-x^2} \, P(x) = \sum_{i=1}^n w_i \, P(x)
\ee
and for $f\in C^{2n}$ there exists $\zeta \in \bR $ such that
\be
\int_{-\infty }^\infty dx\, e^{-x^2} \, f(x) = \sum_{i=1}^n w_i \, f(x)
+\frac{f^{(2n)}(\zeta )}{(2n)!}\, ({\rm H}_n,{\rm H}_n).
\ee
The coefficients $w_i$ are given by
\be
w_i = \frac{2^{n+1} n! \sqrt{\pi}}{[{\rm H}_{n+1}(x_i)]^2} .
\ee
\end{theorem}
Lemma \ref{wickortho} eq.~(\ref{ortho}) yields
\be
<:\Phi^l:_\g \cdot :\Phi^m:_\g \cdot :\Phi^m:_\g >_\g =
C_l^{mn}\, \g^{\frac{l+m+n}{2}}\, l!.
\ee
Thus, using eq.~(\ref{wickherm}), $\g = \frac{1}{2}$,
\be
C_l^{mn} = 2^{-\frac{l+m+n}{2}}\, l!^{-1}\, <{\rm H}_l\cdot
{\rm H}_m\cdot {\rm H}_n>_\g .
\ee
Theorem \ref{GauHerm} implies
\be \label{chst}
C_l^{mn} = \frac{1}{2^{\frac{l+m+n}{2}}\, \sqrt{\pi}\, l!}\,
\int_{-\infty }^\infty d\Phi \, \exp \{ -\Phi^2 \} \,
{\rm H}_l (\Phi ) {\rm H}_m (\Phi ) {\rm H}_n (\Phi ).
\ee
\begin{corollary} \label{Verlinde}
Let  $x_1,x_2,\ldots ,x_N$ be the zeroes of ${\rm H}_n(\Phi )$ and $w_i:=
\frac{2^{N+1} N! \sqrt{\pi}}{[{\rm H}_{N+1}(x_i)]^2}$. Then, for $m+n+l \le
2N-1$, we have
\be \label{AVer}
C_l^{mn} = \frac{1}{2^{\frac{l+m+n}{2}}\, \sqrt{\pi}\, l!}\,
\sum_{i=1}^N w_i\, {\rm H}_l (x_i) {\rm H}_m (x_i) {\rm H}_n (x_i).
\ee
For $l_{max} \in \bN $ and $N\in \bN $, $N\ge \half (3l_{max} +1)$
the equations
\be \label{fixa}
a_l = \sum_{m,n=0}^{l_{max}} \beta^l\, C_l^{mn} a_m a_n, \qquad
l\in \{ 0,\ldots ,l_{max} \}
\ee
imply
\be \label{fixb}
b_j = \sum_{i=1}^N Q_{ji}\, b_i^2,\qquad j\in \{ 1,\ldots ,N\} ,
\ee
where
\be \label{Qeq}
Q_{ji} := \sum_{l=0}^{l_{max}} \frac{2^{-l} \beta^l}{l! \sqrt{\pi}}\,
           w_i\, {\rm H}_l(x_i) {\rm H}_l(x_j)
\ee
and
\be \label{barel}
b_j := \sum_{l=0}^{l_{max}} 2^{-\frac{l}{2}} {\rm H}_l(x_j) a_l .
\ee
\end{corollary}
\vskip5mm\noindent
\underline{\sl Proof:\,} Eq.~(\ref{AVer}) follows immediately from
eq.~(\ref{chst}) and Theorem \ref{GauHerm}. Eq.~(\ref{barel}) and
eq.~(\ref{fixa}) yield
\be \label{Ver1}
b_j = \sum_{l=0}^{l_{max}} 2^{-\frac{l}{2}} {\rm H}_l(x_j)\,
         \sum_{m,n=0}^{l_{max}} \beta^l\, C_l^{mn} a_m a_n.
\ee
Insertion of eq.~(\ref{AVer}) into the rhs of eq.~(\ref{Ver1}) gives
\bea
b_j &:=& \sum_{l=0}^{l_{max}} 2^{-\frac{l}{2}} {\rm H}_l(x_j)\,
          \sum_{m,n=0}^{l_{max}} \beta^l\,
             \sum_{i=1}^N w_i\, {\rm H}_l (x_i) {\rm H}_m (x_i)
                {\rm H}_n (x_i)a_ma_n
\nonu\\ &=&
\sum_{l=0}^{l_{max}} \frac{2^{-l} \beta^l}{l! \sqrt{\pi}}\,
           \sum_{i=1}^N b_i^2 w_i\, H_l(x_i) H_l(x_j)
\nonu\\ &=&
\sum_{i=1}^N Q_{ji}\, b_i^2. \  \  \Box
\eea

\begin{lemma} For $i,j\in \{ 1,\ldots ,N\} $ let $Q_{ij}$ be defined by
eq.~(\ref{Qeq}). Then, we have
\be
\sum_{i=1}^N Q_{ji} = 1,
\ee
for all $j\in \{ 1,\ldots ,N\} $.
\end{lemma}
\vskip5mm\noindent
\underline{\sl Proof:\,} Use Corollary \ref{Verlinde} and the solution
$a_l =\delta_{l,0}$ of eq.~(\ref{fixa}). Insertion of this solution into
eq.~(\ref{barel}) and the use of $H_0(x)=1$ yields $b_j=1$,
for all $j\in \{ 1,\ldots ,N\} $.
Thus eq.~(\ref{fixb}) implies the assertion. $\  \  \Box $

\abstand
The matrix for the linearized RG equation reads
\be
U_{nl} := 2\beta^l \sum_m C_l^{mn} a_m.
\ee
\begin{lemma} \label{Eigtrans}
The eigenvalue problem
\be \label{eigequ}
U(a)u =\lambda u
\ee
is equivalent to
\be  \label{treigequ}
V(b)v = \lambda v,
\ee
where
\be
V_{ji}(b) := 2Q_{ji} b_i, \qquad i,j\in \{ 1,\ldots ,N\}
\ee
and $b$ is related to $a$ by eq.~(\ref{barel}) and correspondingly
$v$ is related to $u$.
\end{lemma}

\begin{lemma} For $j\in \{ 1,\ldots ,N\} $, $m\in \{ 1,\ldots ,l_{max}\} $,
we have
\be
\sum_{i=1}^N Q_{ji} {\rm H}_m(x_i) = \beta^m {\rm H}_m(x_j).
\ee
\end{lemma}

\section{$\epsilon $-Expansion}
\label{EpsExp}
In this section we study the $\epsilon $-expansion for the infrared
fixed points. For dimensions $d_* =\frac{2l_*}{l_*-1}$, $l_*\in
\{ 2,3,\ldots \} $ the ultraviolet fixed point $Z_{UV}$ is a
bifurcation point. In $d_*$ dimensions the UV fixed point
bifurcates into the UV fixed point and the $l_*$-well fixed point
if the dimension is lowered.
We will use the algebra with base $\{u_m,\, m=0,1,2,\ldots \} $ and
multiplication
\be \label{eps0}
u_m\times_\beta u_n = \sum_{l:\atop |m-n| \le l \le m+n} \beta^{2l}
  \cC_l^{mn} u_l,
\ee
where $\beta := 2^{\frac{2-d}{2d}}$.
\begin{lemma} \label{epsfix}
Suppose that $l_*\in \{ 2,3,\ldots \} $ and $d_* :=
\frac{2l_*}{l_*-1}$, $\beta_* := 2^{\frac{2-d_*}{2d_*}}$.
For $d:=d_*-\epsilon $ and $\beta := 2^{\frac{2-d}{2d}}$
consider the $\epsilon $-expansion
\be \label{eps1}
\beta^{-2l} = \beta_*^{-2l} \sum_{m=0}^\infty (\beta^{-2l})^{(m)} \epsilon^m.
\ee
Furthermore, suppose that $z=\sum_{m=0}^\infty z_m u_m $ is a fixed point,
i.~e.
\be \label{eps2}
z=z\times_\beta z.
\ee
This is equivalent to
\be \label{eps3}
\cS_{\beta^{-1}}(z) = z\times_1 z,
\ee
where $\cS_{\beta^{-1}}$ is a linear operator defined by
\be \label{eps4}
\cS_{\beta^{-1}}(u_m) =\beta^{-2m} u_m,
\ee
for all $m\in \bN $. Consider the $\epsilon $-expansion
\be \label{eps5}
z= u_0 + \sum_{k=1}^\infty \epsilon^k \, z^{(k)}.
\ee
Then we have
\be \label{eps6}
z^{(1)} \in \ker [(1-2u_0) \times_{\beta_*}] = \{ \lambda u_{l_*}|\,
  \lambda \in \cC \} =:\cB_{l_*}.
\ee
Let $P_{l_*}:\, \cB \rightarrow \cB $ be the
projection operator defined by $P_{l_*}(\cB ) =
\cB_{l_*}.$ Consider the $\epsilon $-expansion of $\cS_{\beta^{-1}}$
\be \label{eps7}
\cS_{\beta^{-1}} =\sum_{m=0}^\infty \epsilon^m\,
\cS_{\beta_*^{-1}} \circ \cS_{\beta^{-1}}^{(m)} ,
\ee
where the linear operator $\cS_{\beta_*^{-1}}^{(m)} :\, \cB
\rightarrow \cB $
is defined by
\be
\qquad \cS_{\beta^{-1}}^{(m)}(u_l) =
    (\beta^{-2l})^{(m)}\, u_l.
\ee
Then we have
\be \label{eps8}
z^{(1)} =\alpha u_{l_*} , \qquad \alpha :=
     \frac{2(\beta^{-2l_*})^{(1)}}{\cC_{l_*}^{l_*l_*}}
\ee
and
\bea \label{eps10}
(1-P_{l_*})(z^{(k)}) &=& [(1-2u_0)\times_{\beta_*}]^{-1}
   \sum_{m=1}^{k-1}  [
-\cS_{\beta^{-1}}^{(k-m)}((1-P_{l_*})(z^{(m)}))
\nonu\\ &+&
      (1-P_{l_*})(z^{(k-m)}\times_{\beta_*} z^{(m)})],
\eea
and for $k\ge 2$
\bea \label{eps9}
P_{l_*}(z^{(k)}) &=& \frac{1}{(\beta^{-2l_*})^{(1)}} [
   \sum_{m=1}^{k-1} (\beta^{-2l_*})^{(k+1-m)} P_{l_*}(z^{(m)}) -
\nonu\\ &-&
      \sum_{m=2}^{k-1} P_{l_*}(z^{(k+1-m)}\times_{\beta_*} z^{(m)})
  +2P_{l_*}(z^{(1)}\times_{\beta_*} (1-P_{l_*})(z^{(k)}))].
\eea

\end{lemma}
\underline{\sl Proof:\,} The fixed point equation gives
\be \label{eps11}
\sum_{m=0}^\infty \epsilon^m \cS_{\beta^{-1}}^{(m)}
  (u_0 + \sum_{k=1}^\infty \epsilon^k \, z^{(k)}) =
   u_0 + 2u_0\times_{\beta_*} \sum_{k=1}^\infty \epsilon^k \, z^{(k)}+
\sum_{k_1,k_2=1}^\infty \epsilon^{k_1+k_2}z^{(k_1)}
\times_{\beta_*} z^{(k_2)}.
\ee
Since $\cS_{\beta^{-1}}^{(0)} = \eins $ and
\be \label{eps12}
\cS_{\beta^{-1}}^{(m)}(u_0) = \delta_{m,0} u_0,
\ee
we obtain
\be \label{eps13}
(1-2u_0)\times_{\beta_*}z^{(k)} = \sum_{m=1}^{k-1}  (
-\cS_{\beta^{-1}}^{(k-m)}(z^{(m)}) + z^{(k-m)}\times_{\beta_*} z^{(m)}) .
\ee
For $k=1$, we obtain
\be
(1-2u_0) \times_{\beta_*} z^{(1)} = 0.
\ee
This shows eq.~(\ref{eps6}). For $k=2$ we get
\be
(1-2u_0)\times_{\beta_*}z^{(2)} =
-\cS_{\beta^{-1}}^{(1)}(z^{(1)}) + z^{(1)}\times_{\beta_*} z^{(1)}.
\ee
Thus
\be
\cS_{\beta_*^{-1}}^{(1)}(P_{l_*}(z^{(1)})) = P_{l_*}(z^{(1)}
 \times_{\beta_*} z^{(1)}).
\ee
Therefore, $z^{(1)}=\alpha u_{l_*},$
\be
\alpha (\beta^{-2l_*})^{(1)} = \beta^{2l_*} \cC_{l_*}^{l_*l_*} \alpha^2.
\ee
Since $2\beta^{2l_*}=1$ we see eq.~(\ref{eps8}).
Applying the projection operator $P_{l_*}$ to eq.~(\ref{eps13}) we obtain
\be \label{eps14}
0 = \sum_{m=1}^{k-1}  [-\cS_{\beta^{-1}}^{(k-m)}(P_{l_*}(z^{(m)})) +
P_{l_*}(z^{(k-m)}\times_{\beta_*} z^{(m)})] .
\ee
For $k\ge 3$, eq.~(\ref{eps14}) implies
\bea
\lefteqn{-2P_{l_*}(z^{(k-1)}\times_{\beta_*} z^{(1)}) +
    \cS_{\beta^{-1}}^{(1)}(P_{l_*}(z^{(k-1)}))}
\nonu\\ & &
       = -\sum_{m=1}^{k-2}  \cS_{\beta^{-1}}^{(k-m)}(P_{l_*}(z^{(m)})) +
           \sum_{m=2}^{k-2} P_{l_*}(z^{(k-m)}\times_{\beta_*} z^{(m)}).
\eea
Since
\be
\cS_{\beta^{-1}}^{(1)}\circ P_{l_*} = P_{l_*}(z^{(1)}\times_{\beta_*})
   P_{l_*}
\ee
we obtain
\bea \label{eps15}
\cS_{\beta^{-1}}^{(1)}(P_{l_*}(z^{(k-1)}))
       &=& \sum_{m=1}^{k-2}  \cS_{\beta^{-1}}^{(k-m)}(P_{l_*}(z^{(m)})) -
           \sum_{m=2}^{k-2} P_{l_*}(z^{(k-m)}\times_{\beta_*} z^{(m)})
\nonu\\ &+&
2P_{l_*}(z^{(1)}\times_{\beta_*} (1-P_{l_*})(z^{(k-1)})).
\eea
Using $\cS_{\beta^{-1}}^{(1)}(u_{l_*}) =(\beta^{-2l_*})^{(1)}$ and replacing
$k$ by $k+1$ in eq.~(\ref{eps15}) we obtain eq.~(\ref{eps9}).
Applying projection operator $1-P_{l_*}$ to eq.~(\ref{eps13}) we obtain
eq.~(\ref{eps10}). $\  \  \ \Box $
\begin{lemma} \label{lzlk0}
Let $z^{(k)}$ be the coefficients of $\epsilon^k$ in the
$\epsilon $-expansion of a fixed point $z$ in $d_*-\epsilon $ dimensions,
$d_*=\frac{2l_*}{l_*-1}$. Then, we have
\be
z_l^{(k)} =0,\qquad \mbox{if}\  l>l_*\, k.
\ee
\end{lemma}
In the following we will present a bound on the coefficients $z^{(k)}$
of the $\epsilon$-expansion. For the proof of this bound we will need
the following Lemma.
\begin{lemma} \label{lepb1}
Let $(\beta^{-2l})^{(m)}$ be defined by eq.~(\ref{eps1}).
Then we have
\be \label{epb1}
|(\beta^{-2l})^{(m)}| \le \frac{1}{2}\, (\frac{2}{d_*})^m\,
   (2^{\frac{2l}{d_*}}-1).
\ee
Furthermore,
\be \label{epb2}
\sup_{l:\, l\ne l_*} \frac{1}{1-2\beta_*^{2l}} = \frac{1}{1-2^{-
  \frac{1}{l_*}}}.
\ee
\end{lemma}
\underline{Proof:\,} An explicit calculation yields
\be \label{epb3}
(\beta^{-2l})^{(m)} = \sum_{k=0}^{m-1} \frac{1}{(m-k)!} \,
  (-\frac{2l\, \ln 2}{d_*})^{m-k} \, {m-1\choose k}\, d_*^{-m} .
\ee
Using ${m-1\choose k} \le \frac{1}{2} \, 2^m$, we obtain eq.~(\ref{epb1})
By definition of $\beta_*$ we have $2\beta_*^{2l} = 2^{\frac{l_*-l}{l_*}}.$
Therefore, $1-2\beta_*^{2l}$ is minimal for $l=l_*\pm 1.$ Since
\be
2^{\frac{1}{l_*}}-1 > 1-2^{-\frac{1}{l_*}},
\ee
the assertion eq.~(\ref{epb2}) follows. $\  \  \Box $
\begin{lemma} \label{lepsbou}
Let $z^{(k)}$, for all $k\in \bN $, be the coefficients
of the $\epsilon$-expansion. Then there exists constants $K,C$ such that
\be \label{epb4}
|z_l^{(k)}| \le K\, C^k\, (\rho^*_\infty )^{-l} \, \frac{(k-1)!}{l!},
\ee
for all $l,k\in \bN $, where
\be
\rho^*_\infty := \frac{2}{2\beta^2-1}>1.
\ee
\end{lemma}
\underline{Proof:\,} The recursive equations (\ref{eps9}) and
(\ref{eps10}) for the coefficients of the $\epsilon$-expansion
are written in components, for all $l\ne l_*$,
\bea \label{epb6}
z^{(k)}_l &=& [(1-2u_0)\times_{\beta_*}]^{-1}[
   \sum_{m=1}^{k-1}  (
-(\beta^{-2l})^{(k-m)}z^{(m)}_l
\nonu\\ &+&
      (z^{(k-m)}\times_{\beta_*} z^{(m)})_l],
\eea
and
\bea \label{epb5}
z^{(k)}_{l_*} &=& \frac{1}{(\beta^{-2l_*})^{(1)}} [
   \sum_{m=1}^{k-1} (\beta^{-2l_*})^{(k+1-m)} z^{(m)}_{l_*} -
\nonu\\ &-&
      \sum_{m=2}^{k-1} (z^{(k+1-m)}\times_{\beta_*} z^{(m)})_{l_*})
+2P_{l_*}(z^{(1)}\times_{\beta_*}(1-P_{l_*})( z^{(k)}))].
\eea
We proceed by induction. Suppose that eq.~(\ref{epb4}) holds for all
$l^{'}<l.$
Consider the case $l\ne l_*.$
Since $z_l^{(k)}=0$ if $l>l_*\, k$ (see Lemma \ref{lzlk0}), we may suppose
$l\le l_*\, k.$ We want to find a better estimate of
$|(\beta^{-2l})^{(m)}|.$ For that, we distinguish two cases.

a) $l\le \frac{md_*}{2\ln 2}$: By eq.~(\ref{epb3}) follows
\be
|(\beta^{-2l})^{(m)}| \le \frac{1}{2}\, (\frac{2}{d_*})^m\,
\sum_{k=0}^{m-1} \frac{1}{k!} \,
  (\frac{2l\, \ln 2}{d_*})^{k} .
\ee
Thus
\be \label{epb8}
|(\beta^{-2l})^{(m)}| \le \frac{1}{2}\, (\frac{2}{d_*})^m\,
  2^{\frac{m}{\ln 2}} = \frac{1}{2}\, (\frac{2e}{d_*})^m.
\ee
b) $l> \frac{md_*}{2\ln 2}$:  By eq.~(\ref{epb3}) follows
\be \label{epb9}
|(\beta^{-2l})^{(m)}| \le \frac{1}{2}\, (\frac{2}{d_*})^m\,
 \frac{m}{(m-1)!}\, (\frac{2l\, \ln 2}{d_*})^{m-1}
\ee
for all $m\ge 2.$
Define
\be
I := \sum_{m=1}^{k-1} (\beta^{-2l})^{(k-m)}\, K\, C^m\, (m-1)!.
\ee
Then we obtain, using eq.~(\ref{epb9}),
\be
I\le K\, \sum_{m=1}^{k-1} (\frac{2}{d_*})^{k-m}\, \frac{1}{(k-m-2)!}\,
  (\frac{2l\, \ln 2}{d_*})^{k-m-1}\, C^m\, (m-1)! .
\ee
Replacing $m$ by $k-m$ we obtain
\be
I\le K\, \sum_{m=1}^{k-1} (\frac{2}{d_*})^m\, \frac{1}{(m-2)!}\,
  (\frac{2l\, \ln 2}{d_*})^{m-1}\, C^{k-m}\, (k-m-1)! .
\ee
Using $l\le l_* k$ we get
\be
I\le K\, C^{k}\, (k-1)! \, \frac{d_*^2}{4l_*\, \ln 2} \sum_{m=1}^{k-1}
\frac{1}{(m-2)!}\,
  (\frac{4l_*\, \ln 2}{C\, d_*^2})^{m}\, \frac{k^{m-1}}{(k-1)\cdots (k-m)}.
\ee
Therefore
\be \label{epb10}
\sum_{m=1}^{k-1} (\beta^{-2l})^{(k-m)}\, K\, C^m\, (m-1)! \le
  K\, C^k\, (k-1)!\, F(C),
\ee
where $F$ is a fuction obeying
\be
\lim_{C\rightarrow \infty} F(C) =0.
\ee
Thus, eqs.~(\ref{epb6}), (\ref{epb8}) and (\ref{epb10}) imply
\bea \label{epb11}
|z^{(k)}_l|\, \rho^{*l}_\infty \, l! &\le& \frac{1}{1-2^{-\frac{1}{l_*}}}\, [
   \sum_{m=1: \atop m\ge \frac{2l\, \ln 2}{d_*}}^{k-1}  (\frac{1}{2} \,
 (\frac{2e}{d_*})^m
  K\, C^m\, (m-1)!
\nonu\\ &+&
  \, C^k\, (k-1)!\, F(C) + \frac{K^2}{2\beta \sqrt{1-\beta^2}} C^k
  \, (k-m-1)!\, (m-1)! ].
\eea
By eqs.~(\ref{epb11}) follows eq.~(\ref{epb4}) for the case $l\ne l_*.$

\noindent
Consider now the case $l=l_*.$
Using Lemma \ref{lepb1} and Lemma \ref{lVVertaln} of section \ref{NorSca}
we obtain
\bea \label{epb7}
|z^{(k)}_{l_*}| \, \rho_\infty^{*l_*}\, l_*!
&\le & \frac{1}{|(\beta^{-2l_*})^{(1)}|} [
   \sum_{m=1}^{k-1} \frac{1}{4}\, (\frac{2}{d_*})^m\, (2^{l_*-1}-1)
\nonu\\ & &
 K\, C^m\, (m-1)! +
      \sum_{m=2}^{k-1} \frac{K^2}{2\beta \sqrt{1-\beta^2}}\, C^{k+1}\,
        (k-m)!\, (m-1)!
\nonu\\ &+&
2K^2\, C^{k+1} \, (k-1)!].
\eea
By eqs.~(\ref{epb7}) and (\ref{epb11}) follows the assertion. $\  \  \Box $

\section{Convergent Approximation Methods}
\label{BetFlo}
Infrared fixed points are solutions of quadratic equations with an
infinite number of unknowns. For practical calculations a
truncation of these equations to equation with a finite number
of unknowns is necessary.
For a RG analysis it is sufficient to study the truncated
finite-dimensional case. The corresponding error made by truncation
can be estimated. In this section such methods will be
studied.

Firstly, we will explain the beta-function method (cf. \cite{KW91},
\cite{P93}).
By beta-function methods the RG flow of
an infinite number of partition functions is reduced to a
flow of a finite number of coupling
constants. Consider the RG transformation in algebraic representation
\be \label{rgalg}
z^{'} = z\times z.
\ee
Consider a projection operator $P:\, \cB \rightarrow \cB $ such that
the space $P(\cB )$ is finite dimensional. Suppose that an element
$z_0\in P(\cB )$ can be parametrized by $N$ variables $\g_0, \ldots ,
\g_{N-1}$, i.~e.~, $z_0=z_0(\g_0,\ldots ,\g_{N-1}).$ Define two
multiplications $\circ $ and $\bullet $ by
\bea
a\circ b &:=& P(a\times b),
\nonu\\
a\bullet b &:=& (\eins -P)(a\times b).
\eea
Obviously,
\be
a\times b = a\circ b + a\bullet b.
\ee
We split a partition function $z$ into a {\em relevant part}
$z_0 := P(z) \in P(B)$ and an {\em irrelevant part} $r:= (\eins -P)(z) \in
P^c(\cB ) := \cB - P(\cB )$
\be
z = z_0 + r.
\ee
Consider the RG flow of the irrelevant part $r \mapsto r^{'}$
\be \label{irrg}
r^{'} = (z_0 + r) \bullet (z_0 + r).
\ee
For $\hga := (\g_0,\ldots ,\g_{N-1}),$ $z_0 = z_0(\hga )$ we want to define
the fixed point $r_* = r_*(\hga )$ of eq.~(\ref{irrg}), i.~e.,
\be
r_* = z_0 \bullet z_0 + 2z_0 \bullet r_* + r_* \bullet r_* .
\ee
Then, the beta-function $b:\, \bR^N \rightarrow \bR^N $ is defined by
\be
z_0(b(\hga )) = z_0(\hga ) \circ z_0(\hga ) + 2 z_0(\hga )\circ
                r_*(\hga ) + r_*(\hga )\circ r_*(\hga ).
\ee
The fixed points of the RG eq.~(\ref{rgalg}) are determined
by the fixed points $\hga $ of the beta-function $b$, i.~e.,
\be
b(\hga^*) = \hga^* .
\ee
For this fixed point $\hga^*$, we see that
\be
z^* := z_0(\hga^*) + r_*(\hga^*)
\ee
is a fixed point of eq.~(\ref{rgalg}).
\begin{lemma} \label{pnbeta}
For $N\in \bN $ define a projection operator
$P_N:\, \cB \rightarrow \cB $ by
\be
P_N((z_0,z_1,z_2,\ldots )):=
(z_0,z_1,z_2,\ldots ,z_N,0,0,\ldots ).
\ee
Define a function
$F_{z_0} :\, \cB \rightarrow \cB $ by
\be
F_{z_0}(r) := z_0\bullet z_0 + 2z_0\bullet r +r\bullet r,
\ee
for all $z_0 \in P(\cB )$. Define, for all $x\in \cB $, a norm by
\be
\Vert x\Vert := \sum_{n=0}^\infty |x_n|\, \rho^n, \qquad \rho :=
  \frac{\beta }{1-2\beta^2} .
\ee
Let $\kappa >0$ be a positive constant such that
\be
4\kappa \, \sqrt{q_N} <1, \qquad q_N:= \prod_{k=1}^N \frac{2k-1}{2k}
\ee
and define
\be \label{epsdef}
\epsilon (N,\kappa ) =\epsilon := \frac{1}{2\sqrt{q_N}}
  \sum_{n=2}^\infty \frac{1\cdot 3 \cdots (2n-3)}{2^n n!} (4\kappa
   \sqrt{q_N})^n.
\ee
Then, we have
\be \label{subfu}
F_{z_0}(\oU_\epsilon ) \subseteq \oU_\epsilon , \qquad
\oU_\epsilon := \{ x\in \cB |\, \Vert x\Vert \le \epsilon \}
\ee
and $F_{z_0}$ satisfies in $U_\epsilon := \{ x\in \cB |
\, \Vert x\Vert < \epsilon \} $ the Lipschitz condition with
$K:=2\sqrt{q_N} (\kappa +\epsilon )$, i.~e.,
\be \label{lipsch}
\Vert F_{z_0}(r_1) - F_{z_0}(r_2)\Vert < K\, \Vert r_1-r_2\Vert .
\ee
Suppose that $K<1.$ Define recursively
\bea
r^{(0)} &:=& 0
\nonu\\
r^{(n+1)} &:=& F_{z_0}(r^{(n)}).
\eea
Then
\be
\lim_{n\rightarrow \infty } r^{(n)} = \sum_{n=1}^{\infty} (r^{(n)} -
r^{(n-1)}) =r_*
\ee
exists and is a fixed point of $F_{z_0}$, i.~e., $F_{z_0}(r_*) = r_*.$
Furthermore,
\be
\Vert r_*\Vert \le \frac{\sqrt{q_N} \, \kappa^2}{1-K} .
\ee
\end{lemma}
\underline{\sl Proof:\,} By Lemma \ref{lbullbou}
we have for all $r\in \oU_\epsilon $
\be
\Vert F_{z_0}(r)\Vert \le 2\sqrt{q_N} (\kappa + \epsilon )^2 = \epsilon .
\ee
This proves eq.~(\ref{subfu}). Moreover,
\bea
\Vert F_{z_0}(r_1)-F_{z_0}(r_2)\Vert &=&
\Vert 2_0 \bullet (r_1-r_2) + r_1 \bullet (r_1 -r_2) +
  r_2 \bullet (r_1 -r_2) \Vert
\nonu\\ &\le &
2\sqrt{q_N}\, (\kappa +\epsilon )\, \Vert r_1-r_2\Vert .
\eea
This proves the Lipschitz condition eq.~(\ref{lipsch}).
By induction we can prove
\be
\Vert r^{(n)} - r^{(n-1)}\Vert < K^{n-1} \, \sqrt{q_N} \, \kappa^2,
\ee
for all $n\in \bN $, $n\ge 1.$
Thus the series $\sum_{n=1}^{\infty} (r^{(n)} - r^{(n-1)})$ is convergent
and
\be
\Vert r_* \Vert \le \sum_{n=1}^{\infty} \Vert r^{(n)} - r^{(n-1)}\Vert <
  \sum_{n=1}^{\infty} K^{n-1} \, \sqrt{q_N} \, \kappa^2 =
    \frac{\sqrt{q_N} \, \kappa^2}{1-K}.
\ee
This proves the assertion. $\  \  \  \Box $

\abstand
Since $\lim_{N\rightarrow \infty} q_N =0$, the suppositions of Lemma
\ref{pnbeta} can be fulfilled for $N$ large enough.
\begin{corollary}
For $\kappa \in \bR_+$ let $N$ be large enough such that the suppositions
of Lemma \ref{pnbeta} are fulfilled. For $z_0\in P_N(\oU_\kappa )$ let
$r_* \in (1-P_N)(\cB )$ be the solution of $F_{z_0}(r)=r.$ Let $z_0$ be
the solution of
\be
z_0 = z_0 \circ z_0 + 2z_0 \circ r_* +r_* \circ r_*.
\ee
Then $z_* = z_0 +r_*$ is the solution of $z=z\times z$ and
\be
\Vert z_* -z_0\Vert < \frac{\sqrt{q_N}\kappa^2}{1-2\sqrt{q_N} (\kappa +
\epsilon )},
\ee
where $\epsilon $ is defined by eq.~(\ref{epsdef}).
\end{corollary}
\begin{lemma}
Suppose that for $z_0\in P_N(\cB )$ such that
\be
z_0 =z_0 \circ z_0
\ee
the linear operator $(\eins -2z_0)\times $ is invertible. Define
\be
\kappa := \Vert z_0\Vert ,\qquad M:=\Vert O_{z_0}\Vert ,
\ee
where $O_{z_0} := [(\eins -2z_0)\times ]^{-1}.$ Suppose that
\be
4M^2\kappa^2 \sqrt{q_N} <1
\ee
and define
\be
\epsilon := \frac{1}{2M} \, \sum_{n=2}^\infty
  \frac{1\cdot 3 \cdots (2n-3)}{2^n\, n!} \, (4\sqrt{q_N} M^2 \kappa^2)^n.
\ee
Define $G_{z_0}:\, \cB \rightarrow \cB $ by
\be
G_{z_0} (r) := O_{z_0} (z_0 \bullet z_0 +r\times r).
\ee
Then we have
\be \label{gz0sub}
G_{z_0} (\oU_\epsilon ) \subseteq \oU_\epsilon
\ee
and $G_{z_0}$ satisfies on $U_\epsilon$ the Lipschitz condition
with $K:=2M\epsilon .$ Suppose that $K<1.$ Define recursively
\bea
r^{(0)} &:=& 0
\nonu\\
r^{(n+1)} &:=& G_{z_0}(r^{(n)}).
\eea
Then
\be
\lim_{n\rightarrow \infty} r^{(n)} = \sum_{n=1}^\infty (r^{(n)}-r^{(n-1)})
=r_*
\ee
exists and is a fixed point of $G_{z_0}$, i.~e. $G_{z_0} (r_*)=r_*.$
Furthermore,
\be \label{norr*}
\Vert r_*\Vert \le \frac{M\sqrt{q_N}\kappa^2}{1-K}
\ee
and
\be
z=z_0+r_*
\ee
fulfills
\be
z=z\times z.
\ee
\end{lemma}
\underline{\sl Proof:\,} For $r\in \oU_\epsilon $ we have, using Lemma
\ref{lbullbou},
\be
\Vert G_{z_0}(r)\Vert \le M(\sqrt{q_N} \kappa^2 +\epsilon^2) =\epsilon .
\ee
This proves eq.~(\ref{gz0sub}). Since
\bea
\Vert G_{z_0}(r_1)-G_{z_0}(r_2)\Vert  &=& \Vert U_{z_0} (r_* \times
(r_1-r_2) + r_2\times (r_1-r_2)\Vert
\nonu\\ &\le &
2M\epsilon \, \Vert r_1 -r_2\Vert
\eea
we see that $G_{z_0}$ satisfies on $U_\epsilon $ the Lipschitz condition
with $K=2M\epsilon .$
By induction we can prove
\be
\Vert r^{(n)} - r^{(n-1)}\Vert \le K^{n-1} \sqrt{q_N} M \kappa^2,
\ee
for all $n\in \bN ,$ $n\ge 1.$ Thus the series $\sum_{n=1}^\infty (r^{(n)}
-r^{(n-1)})$ is convergent and the bound (\ref{norr*}) follows.
Furthermore, $G_{z_0}(r_*) =r_*$ implies
\be
r_* = 2z_0 \times r_* + z_0 \bullet z_0 + r_* \times r_*.
\ee
Using
\be
z_0\times z_0 = z_0\circ z_0 + z_0\bullet z_0 = z_0 + z_0\bullet z_0
\ee
we obtain
\be
z_0 + r_* = (z_0 + r_*) \times (z_0 + r_*).
\ee
This proves the assertion.  $\  \  \  \Box $

\chapter{Conclusion}
The main conclusion is that the underlying structure of
the renormalization group is that of a nonassociative algebra.
Different presentations of renormalization group transformations
correspond to different choices of bases in this algebra.
For instance the usual form given by Gauss integrable functions
is a special choice. For theories which are not asymptotically
free other choices of bases are better.
A good choice is distinguished by the property that
few parameters suffice to approximate well the renormalization
group flow in a vicinity of fixed points.

The algebra is infinite dimensional. Nontrivial fixed points
turned out to be in the closure of the algebra consisting
of elements of finite degree. The closure is defined
as completion with respect to a certain norm.

In this report we have been concerned with scalar hierarchical
models. Our methods can also be applied to more general
classes of models such as scalar models with many components,
$O(N)$ invariant models, matrix models, gauge theories,
and, we believe, to any quantum field theory.
The first step in the analysis of a general model is
the determination of the constants of multiplication
encoding the algebra structure.

The algebra approach suggests an immediate generalization
of quantum fields which are not perturbations of
Gaussian measures.
For instance we could start from any system of orthogonal
polynomials with respect to any given non-Gaussian measure.

Certain physical aspects have direct counterparts in the
language of the renormalization group algebra.
Subalgebras correspond to universality classes.
Idempotents are fixed points.
Critical indices are given by eigenvalues of the
linearized transformation at a fixed point.
Invariant subspaces under the linearized renormalization
again group correspond to algebra ideals.

In this report we have considered the case
$L^d =2$. More general we could treat the situation when
$L^d$ is a positive integer larger than two.
But physical properties and the general picture of
fixed points and critical behavior are expected to be
independent of the value of $L^d$.

The algebra approach is also useful for computational purposes.
For example we have presented a version of the
$\epsilon$-expansion which can be computed iteratively,
each step consisting of simple algebraic calculations.
Inspecting the Banach algebra we have seen that
the construction of fixed points can be reduced to
a finite dimensional problem.
These considerations lead to convergent expansions
where error terms are controlled using norm estimates.
An example of such a construction is the $\beta$-function
technique.
In terms of special basis we found exact solutions of
the fixed point equation in two dimensions given by
$\Theta$-functions.
A possibility which we have touched upon in the
second chapter is given by calculations with trees.
Tree formulas are useful in various aspects.
They can be used to derive improved convergent expansions.
They can also be used to actually compute for example
fixed points.
Tree formulas are well known in the theory of renormalization.
It seems not to have been realized before that the
reason for their appearance is nonassociativity of
renormalization group algebras.

An important aspect which we have left out is the
construction of Greens functions.
An amusing fact is that they can be computed
exactly at the fixed points.
At the fixed point also operator product expansions
can be calculated exactly and be expressed in terms
of the algebra.
Outside the fixed points Greens functions also serve
to define the critical manifold.
An interesting open problem is the determination of
the critical manifold in terms of the algebra.

We expect our methods to be generalizable to
the full (non-hierarchical) model.
There the main technical problem is to
combine the renormalization group algebra with
polymer algebras and thereby treat the appearance
of nonlocal terms.
An important difference to the hierarchical model is
the wave function term.
Work along these lines is in progress.
\section*{Acknowledgement}
We thank Klaus Pinn and Peter Wittwer for sharing
their insights with us.
Support by the Deutsche Forschungsgemeinschaft is
gratefully acknowledged.
\chapter{Appendices}
\section*{Appendix 1: Fixed Point Equation}
%
We present the fixed point equation
\be
{\it al} = \beta^{2l}\, \sum_{m,n:\atop |m-n|\le l \le
\min (m+n,lmax)} \cC^{mn}_l \, {\it am} \, {\it an},
\ee
for the special case $d=3$, $\beta = 2^{-\frac{d+2}{2d}}$ and $lmax=4$:
\bea
{\it a0}&=&{\it a0}^{2}+2\,{\it a1}^{2}+24\,{\it a2}^{2}+720\,
{\it a3}^{2}+40320
\,{\it a4}^{2}
\\
{\it a1} &=&{\frac {\sqrt [3]{2}{\it a0}\,
{\it a1}}{2}}+\sqrt [3]{2}{\it a1}^{2}+6
\,\sqrt [3]{2}{\it a1}\,{\it a2}+24\,\sqrt [3]{2}{\it a2}^{2}+180\,
\sqrt [3]{2}{\it a2}\,{\it a3}+1080\,\sqrt [3]{2}{\it a3}^{2}
\nonu\\
&+&10080\, \sqrt [3]{2}{\it a3}\,{\it a4}
+80640\,\sqrt [3]{2}{\it a4}^{2}
\\
{\it a2} &=&{\frac {2^{2/3}{\it a0}\,{\it a2}}{8}}
+{\frac {2^{2/3}{\it a1}^{2}}{16
}}+2^{2/3}{\it a1}\,{\it a2}+{\frac {15\,2^{2/3}{\it a1}\,{\it a3}}{4}
}+{\frac {9\,2^{2/3}{\it a2}^{2}}{2}}+60\,2^{2/3}{\it a2}\,{\it a3}
\nonu\\
&+&210\,2^{2/3}{\it a2}\,{\it a4}
+{\frac {675\,2^{2/3}{\it a3}^{2}}{2}}+
5040\,2^{2/3}{\it a3}\,{\it a4}+35280\,2^{2/3}{\it a4}^{2}
\\
{\it a3} &=&{\frac {{\it a0}\,{\it a3}}{16}}
+{\frac {{\it a1}\,{\it a2}}{16}}+{
\frac {3\,{\it a1}\,{\it a3}}{4}}+{\frac {7\,{\it a1}\,{\it a4}}{2}}+{
\frac {{\it a2}^{2}}{2}}+{\frac {45\,{\it a2}\,{\it a3}}{4}}+84\,{\it
a2}\,{\it a4}+75\,{\it a3}^{2}
\nonu\\
&+&1575\,{\it a3}\,{\it a4}
+11760\,{\it a4
}^{2}
\\
{\it a4} &=&{\frac {\sqrt [3]{2}{\it a0}\,
{\it a4}}{64}}+{\frac {\sqrt [3]{2}{\it
a1}\,{\it a3}}{64}}+{\frac {\sqrt [3]{2}{\it a1}\,{\it a4}}{4}}+{
\frac {\sqrt [3]{2}{\it a2}^{2}}{128}}+{\frac {3\,\sqrt [3]{2}{\it a2}
\,{\it a3}}{8}}+{\frac {21\,\sqrt [3]{2}{\it a2}\,{\it a4}}{4}}
\nonu\\
&+&{ \frac {225\,\sqrt [3]{2}{\it a3}^{2}}{64}}
+105\,\sqrt [3]{2}{\it a3}\, {\it a4}
+{\frac {3675\,\sqrt [3]{2}{\it a4}^{2}}{4}}
\eea
\newpage
\section*{Appendix 2: Constants of Multiplication}
%
We present a table for all constants of multiplication $B(l,m,n)$ for
$l,m,n \le 4$. Coefficients which do not occur in the table are either
permutations of the arguments of $B$ or are equal to zero.
\begin{table}[h]
\small
\begin{center}
\begin{tabular}{|r|c|c|r|}
\hline
  \mc{1}{|c|}{$l$}  &
  \mc{1}{c|}{$m$} &
  \mc{1}{c|}{$n$} &
  \mc{1}{c|}{$B(l,m,n)$}\\
\hline
  0 & 0 & 0  & 1=1.00000000 \\
  1 & 1 & 0  & 1=1.00000000 \\
  1 & 1 & 1  & $2*\sqrt{2}=2.82842712$ \\
  2 & 1 & 1  & $\sqrt{6}=2.44948974$ \\
  2 & 2 & 0  & 1=1.00000000 \\
  2 & 2 & 1  & $4*\sqrt{2}=5.65685425$ \\
  2 & 2 & 2  & $6*\sqrt{6}=14.69693846$ \\
  3 & 2 & 1  & $\sqrt{15}=3.87298335$ \\
  3 & 2 & 2  & $8*\sqrt{5}=17.88854382$ \\
  3 & 3 & 0  & 1=1.00000000 \\
  3 & 3 & 1  & $6*\sqrt{2}=8.48528137$ \\
  3 & 3 & 2  & $15*\sqrt{6}=36.74234615$ \\
  3 & 3 & 3  & $40*\sqrt{5}=89.44271912$ \\
  4 & 2 & 2  & $\sqrt{70}=8.36660027$ \\
  4 & 3 & 1  & $2*\sqrt{7}=5.29150262$ \\
  4 & 3 & 2  & $8*\sqrt{21}=36.66060557$ \\
  4 & 3 & 3  & $15*\sqrt{70}=125.49900400$ \\
  4 & 4 & 0  & 1=1.00000000 \\
  4 & 4 & 1  & $8*\sqrt{2}=11.31370850$ \\
  4 & 4 & 2  & $28*\sqrt{6}=68.58571280$ \\
  4 & 4 & 3  & $112*\sqrt{5}=250.43961350$ \\
  4 & 4 & 4  & $70*\sqrt{70}=585.66201860$ \\
\hline
 \end{tabular}
  \parbox[t]{.85\textwidth}
  {
  \caption[Coefficients $B(l,m,n)$ for $l,m,n \le 4$.]{\label{demotab}
  The coefficients $B(l,m,n)$ for $l,m,n \le 4$.}
  }
\end{center}
\end{table}

\end{document}